\documentclass{aa}
\usepackage{graphicx}
\usepackage{txfonts}
\usepackage{mathrsfs}
\usepackage{amssymb,amsmath}
\usepackage{newtxtext,newtxmath}
\usepackage{amsmath}
\usepackage{amssymb}
\DeclareMathAlphabet{\mathbi}{OT1}{ptm}{bx}{it}
\SetMathAlphabet\mathbi{bold}{OT1}{ptm}{bx}{it}

\usepackage{lipsum}
\usepackage{stfloats}
\usepackage{booktabs}
\usepackage{comment}

\usepackage[english]{babel}
\usepackage[colorinlistoftodos,prependcaption,textsize=tiny]{todonotes}

\usepackage{hyperref}
\hypersetup{
    colorlinks=true,
    citecolor=blue,
    linkcolor=blue,
    filecolor=magenta,      
    urlcolor=cyan,
}

\usepackage{subfig}

\newcommand{\appropto}{\mathrel{\vcenter{
  \offinterlineskip\halign{\hfil$##$\cr
    \propto\cr\noalign{\kern2pt}\sim\cr\noalign{\kern-2pt}}}}}
    
\begin{document} 

   \title{A simulation-based quality-control framework for the broad-line region radius-luminosity relation\thanks{The code can be downloaded from: \url{https://github.com/Juri-W-S/pyRMTools.git}. \fnmsep \thanks{Instructions and specific examples used in this paper can be found in: \url{https://github.com/Juri-W-S/pyRMTools}}}}
   \titlerunning{A quality-control framework for the $R$-$L$ relation}

   \author{J. W. Seib\inst{1}\href{https://orcid.org/0009-0007-7791-4002}{\includegraphics[scale=0.5]{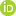}}
          \and
          F. Pozo Nu\~nez\inst{2}\href{https://orcid.org/0000-0002-6716-4179}{\includegraphics[scale=0.5]{orcidicon.png}}
          \and
          S. E. I. Bosman\inst{1,3}\href{https://orcid.org/0000-0001-8582-7012}{\includegraphics[scale=0.5]{orcidicon.png}}
           }
    
   \institute{Institute for Theoretical Physics, Heidelberg University, Philosophenweg 12, D–69120, Heidelberg, Germany \\
              \email{seib@thphys.uni-heidelberg.de}
              \and
   Astroinformatics, Heidelberg Institute for Theoretical Studies, Schloss-Wolfsbrunnenweg 35, 69118 Heidelberg, Germany\\
   \email{francisco.pozon@gmail.com}\and Max-Planck-Institut für Astronomie, Königstuhl 17, 69117 Heidelberg, Germany}

   \date{Received 20 July, 2026; accepted 20 August, 2026}

\abstract
{The broad-line region (BLR) radius-luminosity ($R$-$L$) relation
underpins single-epoch black hole mass estimates in active galactic
nuclei (AGNs). The published H$\beta$ sample is heterogeneous in terms of
observational quality and it remains unclear how much of its scatter
is intrinsic, rather than caused by systematics in lag recovery.}
{We aim to quantify the contribution of unreliable lag recovery to
the observed scatter of the H$\beta$ $R$-$L$ relation and to provide
the community with the tools required for such quality control measures.}
{We compiled a publicly available database of reverberation mapping
(RM) measurements for $\sim$1200 AGNs from 32 campaigns spanning more
than three decades, storing lags, luminosities, line widths, black
hole masses, and observational metadata. On this basis, we developed a
simulation-based consistency framework: damped random-walk light
curves were sampled according to each campaign's reported baseline,
cadence, and signal-to-noise ratio, while the lags were recovered with the
interpolated cross-correlation function. A comparison of the expected,
reported, and simulation-retrieved lags defined the four-tier flagging
scheme. The $R$-$L$ relation was refit with the Bayesian
nested-sampling code UltraNest for progressively cleaner subsamples,
while the dependence on the assumed reference relation was removed by
constructing consensus samples across three reference slopes.}
{Although some model-dependent biases could persist, of the 248 H$\beta$ sources, $\approx40\%$ of them display a discrepancy between the reported and simulation-retrieved lags, while $\approx5\%$ exhibit a direct
inconsistency between the expected and retrieved lags. Excluding
flagged sources and correcting for model bias reduces the inferred
intrinsic scatter from $\sigma = 0.26^{+0.02}_{-0.01}$~dex to
$\sigma = 0.11\pm0.01$~dex, with a model-bias-corrected slope of
$\alpha = 0.48\pm0.02$. This is consistent with the photoionisation
expectation of $\alpha = 0.5$.}
{A significant fraction of the observed scatter in the current
H$\beta$ $R$-$L$ relation arises from observational limitations and
lag-recovery biases, rather than from intrinsic AGN diversity. The
database, simulation code, and the campaign-planning tool RM-Scout
have been made publicly available.}

   \keywords{galaxies: active
          --galaxies: nuclei --quasars: general
          --quasars: supermassive black holes --galaxies: Seyfert
               }

   \maketitle

\section{Introduction}
\label{sec:Intro}
Active galactic nuclei (AGNs) play a central role in galaxy evolution
across cosmic time \citep[e.g.][]{Fan2023}.
The masses of the supermassive black holes (SMBHs) that power AGNs
represent key observables linking nuclear activity to the surrounding
galaxy \citep{Capelo2023}, as evidenced by tight scaling relations between SMBH mass and
host-galaxy properties, such as the stellar velocity dispersion
\citep[e.g.][]{Ferrarese2000}.
Black holes with $M_{\rm BH}\sim10^9\,M_\odot$
have been detected at redshifts $z>6$, challenging current models of
black hole growth \citep{Inayoshi2020}.
Assessing whether this tension is physical or partly an artifact of mass estimation requires single-epoch (SE) calibrations whose systematics are well understood.

Measuring SMBH masses at high redshift relies heavily on
SE virial mass estimators calibrated via
reverberation mapping (RM; e.g. \citealt{Shen2013}).
In an RM campaign, the time lag, $\tau$, between variability in the
AGN ionising continuum and the response of a broad emission line
is measured, yielding the size, $R = c\tau$, of the broad-line region
(BLR; \citealt{Blandford1982}).
Combining this parameter with the velocity width, $W$, of the emission line
gives the virial mass via
\begin{equation}
  \label{eq:virial}
  M_{\rm BH} = f \,\frac{R\,W^2}{G},
\end{equation}
where $f$ is the dimensionless virial factor encoding the unknown BLR
geometry and kinematics. This RM-based framework also provides the foundation for SE virial mass estimates. In particular, \cite{Kaspi2000} empirically calibrated the BLR radius against the AGN optical luminosity, establishing one of the first widely used radius-luminosity ($R$-$L$) relations for the H$\beta$ BLR over a broad luminosity range. 
This calibration makes it possible to infer the BLR size from the continuum luminosity, thereby bypassing dedicated RM campaigns and estimating the SMBH mass through the virial relation from a SE spectrum.
The relation takes the form of
\begin{equation}
  \label{eq:rl}
  \log\left(\frac{R}{\text{lt-day}}\right) =
  \beta + \alpha\,\log\left(\frac{L_{5100}}{10^{44}\,\text{erg\,s}^{-1}}\right)
  + \epsilon,
\end{equation}
where $\alpha$ is the slope, $\beta$ the normalisation, and
$\epsilon$ represents intrinsic scatter with a standard deviation,
$\sigma$.
 
From the photoionisation arguments, a slope of $\alpha = 0.5$ would be
expected \citep{Osterbrock}; however, fits to the observed H$\beta$ sample yield slopes in the
range $\alpha \approx 0.4$-$0.7$ with intrinsic scatter of
$\sigma \approx 0.13$-$0.3$\,dex
\citep[e.g.][]{Kaspi2000,Bentz2013,Shen2024}.
This casts doubt on the accuracy of SE mass estimates,
particularly when the relation is extrapolated to the high-luminosity,
high-redshift regime where an independent RM calibration is scarce.

A persistent challenge is that published RM samples are heterogeneous
in terms of observational quality: campaigns differ in baseline duration,
observing cadence, seasonal coverage, and signal-to-noise ratio
(S/N). The most widely used lag-recovery algorithm, the interpolated
cross-correlation function \citep[ICCF;][]{Gaskell1986,Peterson1993},
is sensitive to these parameters. However, its performance under realistic
campaign conditions has not been well characterised for the full published
sample.
 
Alternative approaches, such as \texttt{JAVELIN} 
\citep{Zu2011}, \texttt{CREAM} \citep{Starkey2016},
and \texttt{GPCC} \citep{Pozo2023}, incorporate stochastic variability
models and transfer functions and can improve lag recovery for sparsely
sampled or noisy light curves \citep{Li2019}. However, these approaches
also introduce additional assumptions about the continuum variability
and BLR response. Since our aim is to assess the consistency of published
lags rather than to derive new model-dependent lag estimates, we adopted
the ICCF as a common, minimally model-dependent recovery method. We sought to
explicitly characterise its performance under the heterogeneous
observational conditions represented in the published sample.

There are two main barriers to addressing this problem.
First, the relevant observational metadata are scattered across
dozens of publications with no common data standard.
Second, while individual surveys have used recovery simulations to assign internal quality flags (e.g. \citealt{Penton2022,Malik2023}), no such framework has been applied uniformly across the heterogeneous published sample.

In this paper we address both issues.
We present (i) a publicly available database of RM
measurements covering $\sim$1200 AGNs from 32 campaigns spanning
more than three decades, (ii) a simulation-based consistency
framework that applies the ICCF to damped random walk (DRW) light curves sampled according
to each campaign's actual observational parameters, and (iii) an application to the H$\beta$ $R$-$L$ sample yielding a bias-corrected calibration and a revised SE mass estimator.

The paper is structured as follows.
In Sect.~\ref{sec:database} we describe the database content,
architecture, and Python interface.
In Sect.~\ref{sec:simulations} we present the simulation framework,
including the ICCF implementation, bias characterisation, and
parameter exploration.
In Sect.~\ref{sec:rl_analysis}, we describe how we applied the framework to the
H$\beta$ $R$-$L$ sample, define the flagging scheme, and present
the refitted $R$-$L$ parameters.
We discuss implications in Sect.~\ref{sec:discussion} and summarise
our conclusions in Sect.~\ref{sec:conclusions}.

\section{The reverberation mapping database and software package}
\label{sec:database}

To facilitate the use of the database and to provide a reproducible
interface for the analyses presented in this work, we developed the
open-source Python package \texttt{pyRMTools}. The package consists
of two branches, the first giving access to the reverberation mapping
database including utilities for structured querying and scientific use,
and the second making the simulation framework developed in 
Sect.~\ref{sec:simulations} available as a planning tool for future
RM campaigns. The database can be distributed either as a stand-alone
JSON archive or hosted through a MongoDB\footnote{\url{https://www.mongodb.com}}
back-end, both options exposing an identical Python interface.

\subsection{Scope and data sources}
\label{sec:db_scope}

The database compiles RM results from 32 campaigns published between
1994 and 2026, containing data for approximately 1200 individual AGNs.
The campaigns are listed in Table~\ref{tab:campaigns}.
The sample spans nearly three decades of RM observations and includes
the landmark long-baseline programs (e.g.\ \citealt{Kaspi2000}),
the high-cadence AGN STORM campaign \citep{Denney2010}, the RM
compilation with luminosity corrections \citep{Bentz2013}, and the most 
recent large-scale programs such as SDSS-V RM \citep{Shen2024}.

\begin{figure}
  \centering
  \includegraphics[width=\linewidth]{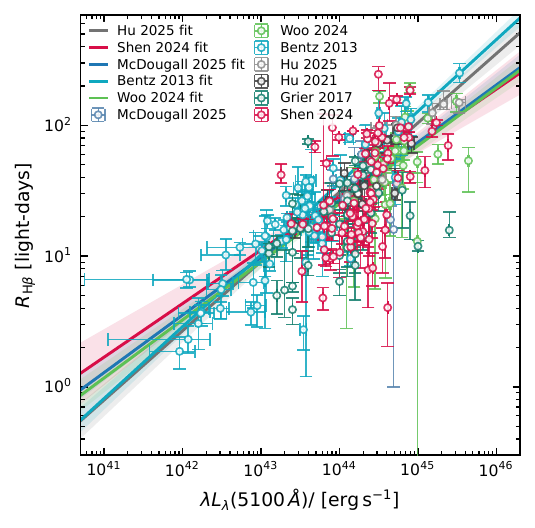}
  \caption{Current state of the H$\beta$ BLR radius-luminosity
           relation compiled from the literature
           (\citealt{Bentz2013,Grier2017,Hu2021,Shen2024,Woo2024,Hu2025,McDougall2025}; colours as in the legend).
           Solid lines show the best-fit relations from each study
           with shaded areas indicating their uncertainties;
           The compilation spans nearly four orders of magnitude in luminosity and comprises more than 200 measurements of 186 AGNs.}
  \label{fig:rl_current}
\end{figure}

The current state of the H$\beta$ $R$-$L$ relation, compiled from
the literature, is shown in Fig.~\ref{fig:rl_current}.
The relation spans nearly four orders of magnitude in luminosity, yet
a large fraction of sources lie systematically below the fitted
relations, motivating the quality-control framework developed in this
work.

\begin{table}[htbp]
  \caption{Reverberation mapping campaigns included in the database.}
  \label{tab:campaigns}
  \centering
  \small
  \begin{tabular}{ll}
    \hline\hline
    Reference & Program / Notes \\
    \hline
    \cite{Stirpe1994}       & Early single-object campaigns \\
    \cite{Winge1996}        & \\
    \cite{Santos-Lleo1997} & \\
    \cite{Dietrich1998}     & \\
    \cite{Peterson1998}     & \\
    \cite{Kaspi2000}        & PG quasar sample \\
    \cite{Santos-Lleo2001} & \\
    \cite{Peterson2002}     & \\
    \cite{Peterson2005}     & \\
    \cite{Bentz2006a}        & \\
    \cite{Denney2006}       & \\
    \cite{Metzroth2006}     & \\
    \cite{Bentz2007}        & \\
    \cite{Denney2009}       & \\
    \cite{Bentz2009b}       & LAMP \\
    \cite{Denney2010}       & MDM-based H$\beta$ campaign \\
    \cite{Grier2012}        & MDM-based H$\beta$ campaign \\
    \cite{Dietrich2012}     & \\
    \cite{Bentz2013}        & H$\beta$ RM compilation \\
    \cite{Peterson2014}     & \\
    \cite{DeRosa2015}      & AGN STORM \\
    \cite{Grier2017}        & SDSS-RM pilot \\
    \cite{Lira2018}         & C\,{\sc iv} sample \\
    \cite{Hoormann2019}     & OzDES RM\\
    \cite{Hu2021}           & \\
    \cite{Kaspi2021}        & two-decade high-luminosity campaign\\
    \cite{Woo2024}          & SNU-AGN monitoring project\\
    \cite{Shen2024}         & SDSS-V RM \\
    \cite{Hu2025}           & \\
    \cite{Penton2025}       & OzDES RM\\
    \cite{McDougall2025}    & OzDES RM\\
    \cite{Bai2026}          & \\
    \hline
  \end{tabular}
\end{table}

The database does not claim completeness. The focus is on campaigns providing multi-object samples with sufficient metadata for population-level analyses; a small number of landmark single-object campaigns are additionally included. Notes and a README file accompany the database to document the
heterogeneous assumptions adopted by different authors (cosmological
parameters, host-galaxy subtraction methods, extinction corrections,
and calibration choices).

\subsection{Database structure}

Each AGN is represented by a dedicated object. Individual observables
are organised into measurement collections (e.g. lags, luminosities, masses),
where every measurement contains its associated metadata such as publication
source, notes on assumptions, uncertainties and observational setup.
In addition to the object-oriented representation, a publication view is available — for the whole database or for individual objects — containing all measurements from a specified publication.
In Appendix~\ref{app:pyRMTools} we describe the class structure and API syntax in more detail.

\subsection{Installation and hosting}

The package can be installed directly from GitHub.
By default, the distributed JSON version of the database is loaded, making the package usable directly after installation. 
Users experienced with databases may instead connect to a locally hosted MongoDB back-end without any change to the interface. 
This option allows for MongoDB-style queries via the Python package \texttt{pymongo}, and the MongoDB Compass graphical client additionally allows for a visual inspection of the collection without needing to write code.

\subsection{Python utility functions}

To demonstrate the database interface, we show how to access data for an individual AGN. The database is loaded via

\begin{verbatim}
import pyRMTools as qrm
db = qrm.Database.from_json()
\end{verbatim}

and querying a single AGN is done via

\begin{verbatim}
agn = db.get('PG 0052+251').
\end{verbatim}

Measurements are stored as properties of the AGN. Unique quantities (such as the position) are stored as single measurements, whereas quantities that may differ between publications are stored as collections of measurements. This causes a slight difference in syntax, as accessing values from a collection requires iterating over the list. Next, 

\begin{verbatim}
pos = agn.position
print('RA, DEC', pos.value, pos.unit)
\end{verbatim}

prints the position and its unit and

\begin{verbatim}
luminosities = agn.luminosity(5100)
for lum in luminosities:
    print(lum.value, lum.error, 
        lum.unit, 
        qrm.reference_finder(lum.source))
\end{verbatim}

\noindent prints all individual $\lambda L_\lambda\left(5100\AA\right)$ measurements 
including uncertainties and the publication the measurement originated from. 

All values in the database assume the cosmologies of the publication that derived the value.
Luminosities computed under a publication's assumed cosmology can be converted to a user-chosen cosmology via the \texttt{astropy.cosmology} module.

\begin{verbatim}
from astropy.cosmology import LambdaCDM
import astropy.units as u
new_cosmology = LambdaCDM(H0 = 67*u.km/u.s/u.Mpc,
                        Om0 = 0.32,Ode0 = 0.68)
for lum in luminosities:
    print('Old:', lum.value, lum.cosmology)
    lum.convert(new_cosmology)
    print('New:', lum.value)
\end{verbatim}

\subsection{RM preparation tool}

Based on our simulation framework (see Sect.~\ref{sec:simulations})
we created a tool that estimates the likelihood of a successful and unbiased lag recovery for given observing parameters, intended for the planning stage of a RM campaign.

The simulation is called up via
\begin{verbatim}
result = qrm.scout(luminosity = 1e43, z = 0, 
                    baseline = 300, cadence = 4, 
                    sn = 100)
\end{verbatim}

and returns the \texttt{result} class. The recovered lag and its uncertainties (16th/84th percentiles of the recovery distribution) are accessed via

\begin{verbatim}
print('lag', '+', '-')
print(result.lag, result.error_plus, 
            result.error_minus)
\end{verbatim}

Diagnostic plots, such as the bias histogram
(analogous to Fig.~\ref{fig:iccf_bias}) can be viewed by calling up

\begin{verbatim}
result.plot.bias_histogram().
\end{verbatim}

The functionality and interpretation of the \texttt{scout} module are described in Appendix~\ref{app:RM-Scout}.

\section{Simulation framework}
\label{sec:simulations}

\subsection{Motivation and overview}
\label{sec:sim_motivation}
 
A central challenge in using published RM lags to calibrate the
$R$-$L$ relation is that some lags may be poorly constrained by their
observational campaigns.
Sparse cadence, short baselines, large seasonal gaps, and low S/N can
all cause the ICCF to fail to recover the true lag, or to recover a
biased estimate.
Our simulation framework is designed to quantify this risk for each
published measurement using its own reported observational parameters.
 
The approach comprises three steps: (i) generate realistic AGN continuum
and emission-line light curves; (ii) sample them according to the
campaign's observational parameters; and (iii) apply the ICCF and compare
the recovered lag to the known input lag. Since the purpose of the simulations is to test whether the published lag could have been recovered under the reported observing conditions, the choice of lag-recovery method
is central to the interpretation.

As discussed in Sect.~\ref{sec:Intro}, our goal is not to derive a new best-fit lag, but to
perform a consistent and conservative assessment of published
measurements. We therefore adopted the ICCF throughout this work, which enabled
us to apply the same recovery procedure to all simulated campaigns
without introducing a specific variability or BLR model.

\subsection{Light curve generation}
\label{sec:lc_generation}
 
The AGN continuum variability can be well described by a damped random walk \citep[DRW;][]{MacLeod2010}, whose power spectral density follows a power law with an index of $\beta_{\rm DRW} = 2$ at high frequencies.
We can generate DRW continuum light curves using the \texttt{stingray}
Python package \citep{stingray} with the signal mean set to 1.
The emission-line light curve is then obtained by convolving the
continuum with a normalised top-hat transfer function,
\begin{equation}
  \Pi(t) =
  \begin{cases}
    1 & \text{if } \tfrac{1}{2}\tau < t < \tfrac{3}{2}\tau, \\
    0 & \text{otherwise,}
  \end{cases}
\end{equation}
centred on the input lag $\tau$ with a width equal to half the lag.
This is a simple but standard approximation to the BLR transfer
function \citep[e.g.][]{Zu2011}.
 
The variability amplitude (rms) appropriate to each object is
estimated using the structure-function scaling relation of
\citet{Morganson2014}:
\begin{align}
  \label{eq:sf}
  \mathrm{SF} = 0.079 &
  \left(1+z\right)^{0.15}
  \left(\frac{L_{5100}}{10^{46}\,\mathrm{erg\,s^{-1}}}\right)^{-0.2}
  \notag \\
  & \times
  \left(\frac{510\,\mathrm{nm}}{1000\,\mathrm{nm}}\right)^{-0.44}
  \left(\frac{T}{365.25\,\mathrm{d}}\right)^{0.246},
\end{align}
with $\mathrm{rms} = \mathrm{SF}/\sqrt{2}$ and $T$ set to the
campaign baseline.

\subsection{Observational sampling}
\label{sec:obs_sampling}
 
The generated light curves were down-sampled to mimic the observational
cadence and coverage of the corresponding RM campaign.
An evenly spaced observing grid was constructed from the cadence
$\delta T$ and total baseline $T$.
Seasonal gaps of approximately 180 days were inserted every 365 days
to approximate the annual visibility window of a ground-based target.
We note that the true window depends on the declination of the source and the latitude of the observing site; therefore, it varies between campaigns. 
We nevertheless adopted a single prescription for all objects: the realised gaps in published campaigns are set by scheduling, weather, and time allocation as much as by geometric visibility. In addition, several programmes combine sites at different latitudes, so an object-specific visibility model would not reproduce the actual sampling. A uniform, conservative gap therefore allows us to avoid introducing object-dependent assumptions into the flagging, while the resulting effect on each simulation is reported through $f_{\rm sim.~cover}$ and $N_{\rm sim}$ in Table~\ref{tab:sim_summary}. Gaussian white noise was then added to the sampled light curves with
a standard deviation set by the adopted S/N.
 
When the cadence was not explicitly reported in the literature, it was
estimated from the total number of epochs, $N$, and the effective
observable baseline,
\begin{equation}
  \delta T = \frac{T_{\mathrm{eff}}}{N_{\mathrm{epochs}}}.
\end{equation}
When the S/N was not reported, we either estimated it from the mean flux-to-uncertainty ratio of the published light curve or (when no light curve was available) we adopted a default of S/N~$= 100$.
These simplifications mean that the simulated observational patterns are approximations; the effects of weather-related gaps, individual epoch failures, and varying photometric conditions are not modelled. This is coupled with the fact that seasonal gaps not necessarily follow our exact assumptions.
Despite these limitations, the framework can simulate
$\simeq 96$\% of the H$\beta$ $R$-$L$ sample. The remaining $\approx4\%$ are unable to be simulated by the framework, due to unavailability of essential observational metadata, such as observational baseline and observed epochs.

\begin{table}[t]
  \caption{Median ICCF bias, $b$, under ideal observing conditions.}
  \label{tab:iccf_bias}
  \centering
  \begin{tabular}{lccc}
    \hline\hline
    $t_{\rm unit}$ & $r = 0.5\,r_{\rm max}$ &
                     $r = 0.6\,r_{\rm max}$ &
                     $r = 0.8\,r_{\rm max}$ \\
    \hline
    $0.8\,\delta T$ & ---   & $-0.71\%$ & $-0.25\%$ \\
    $\delta T$      & $-1.18\%$ & $-1.10\%$ & $-0.35\%$ \\
    $1.2\,\delta T$ & ---   & ---   & $-0.31\%$ \\
    \hline
  \end{tabular}
  \tablefoot{Entries marked '---'\ indicate parameter combinations
  where one setting already produced a strongly biased result and the
  combination was not investigated further.}
\end{table}

\subsection{ICCF implementation and parameter optimisation}
\label{sec:iccf}
 
The ICCF was computed using functions from the \texttt{pyCCF} package
\citep{Sun2018}. The ICCF centroid, $\tau_{\rm cent}$, was computed from the portion of
the cross-correlation function that exceeds a fraction, $r/r_{\rm max}$,
of its peak value, while the lag grid spacing is denoted as $t_{\rm unit}$.
 
\begin{figure}
    \centering
    \includegraphics[width=1\linewidth]{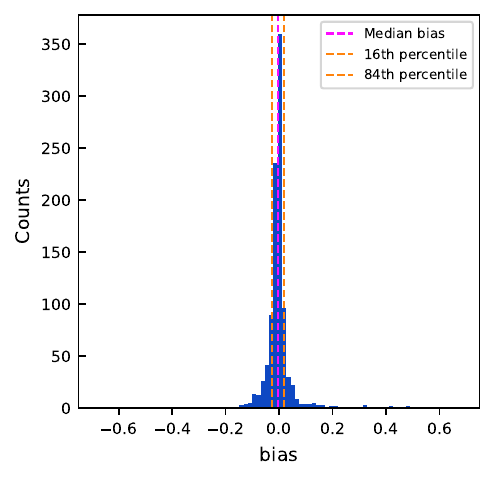}
    \caption{Bias histogram under ideal observing conditions using 
    $t_{\rm unit}=0.8\cdot\delta T$ and $r = 0.8\,r_{\rm max}$.}
    \label{fig:iccf_bias}
\end{figure}

Before applying the ICCF to real campaign simulations, we sought to characterise
its performance under ideal conditions. We generated 1000 DRW light-curve pairs with a known input lag of
50 days, sampled them with an hourly cadence over a long baseline,
add no noise, and applied the ICCF. We defined the bias as
\begin{equation}
\label{eq:bias}
  b = \frac{\tau_{\rm recovered}}{\tau_{\rm true}} - 1
\end{equation}
and took the median over the 1000 realisations.
We explored a grid of parameters:
$r/r_{\rm max} \in \{0.5, 0.6, 0.8\}$ and
$t_{\rm unit} \in \{0.8\,\delta T,\, \delta T,\, 1.2\,\delta T\}$.
Table~\ref{tab:iccf_bias} summarises the median bias for each
combination, and Fig.~\ref{fig:iccf_bias} shows the bias histogram 
for the least biased setup.

The ICCF consistently underestimates the true lag, but at a negligible level for all tested configurations.
The combination $t_{\rm unit} = 0.8\,\delta T$ and
$r = 0.8\,r_{\rm max}$ yields the smallest bias of $-0.25\%$, and we
adopted this configuration for all subsequent simulations.
We verified the robustness of this result by repeating the test with a
realistic variability amplitude of rms~$= 0.1$ ($b = -0.29\%$) and
with an input lag of 100 days ($b = -0.31\%$). Therefore, we follow the assumption that any additionally introduced bias is a consequence of the observing conditions. 

\subsection{Simulation parameter study}
\label{sec:sim_parameterstudy}
 
To understand how the ICCF performance depends on observational
parameters, we systematically varied the baseline, $T$, cadence,
$\delta T$, and S/N around a fiducial input lag of $\tau = 50$~days.
We note that all time parameters are expressed in units of the input lag for
generality. 
 
We considered three simulation set-ups of increasing realism:
\begin{itemize}
  \item \textit{Trial 1}: uniformly sampled light curves with seasonal gaps, with the lag search range defined from the nominal baseline, $T$.
  \item \textit{Trial 2}: same as trial 1, but with the lag search range defined from the actual light-curve length.
  \item \textit{Trial 3}: same as trial 2, with an additional detrending step applied before the cross-correlation.
\end{itemize}
 
\begin{figure*}
    \centering
    \includegraphics[width=\linewidth]{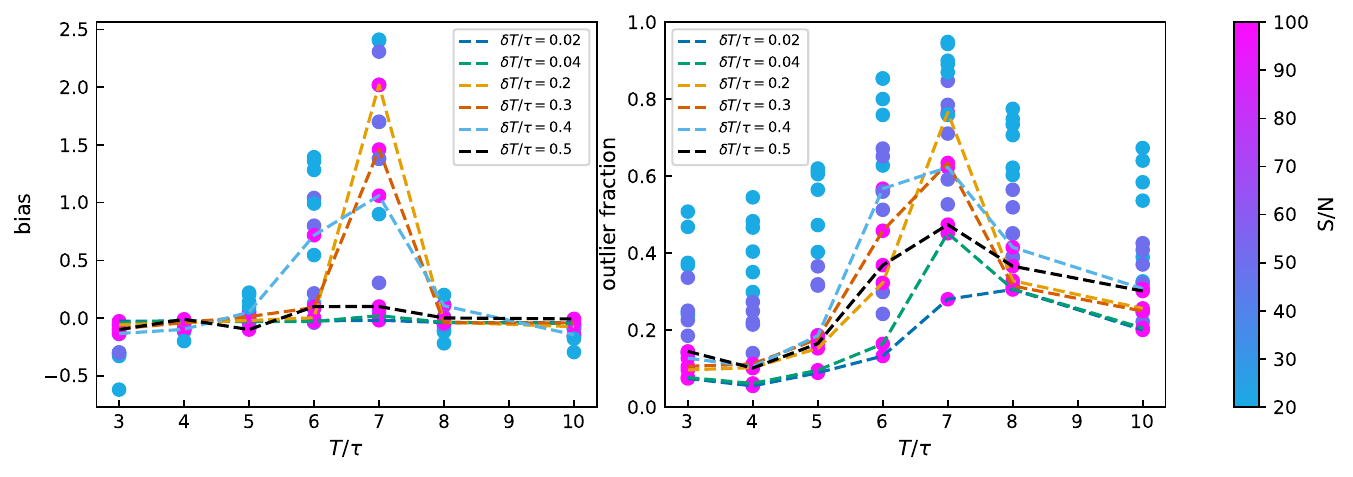}
    \includegraphics[width=\linewidth]{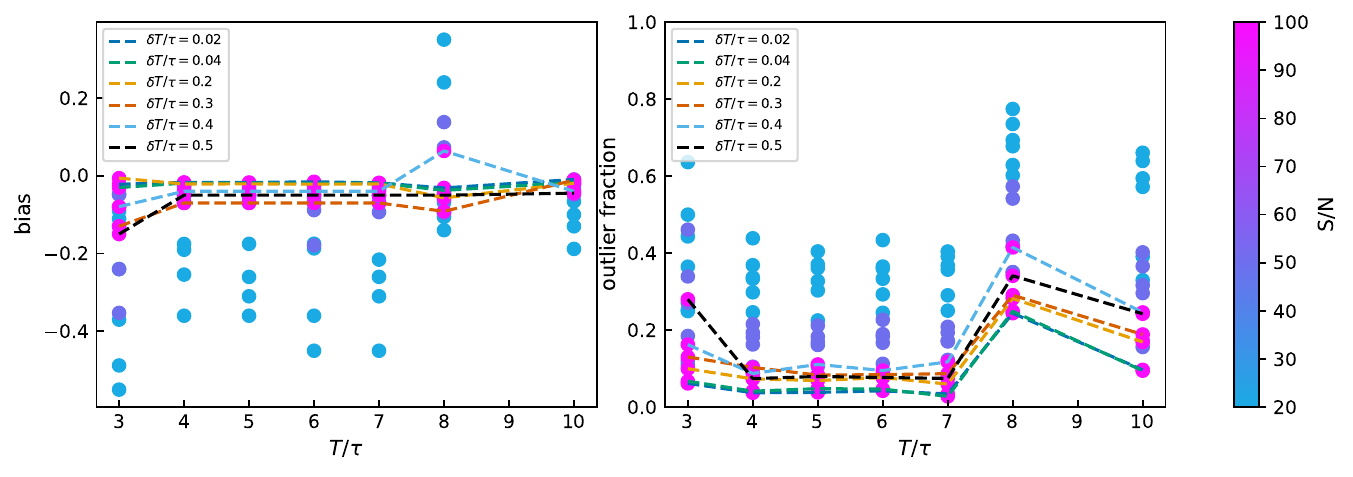}
    \includegraphics[width=\linewidth]{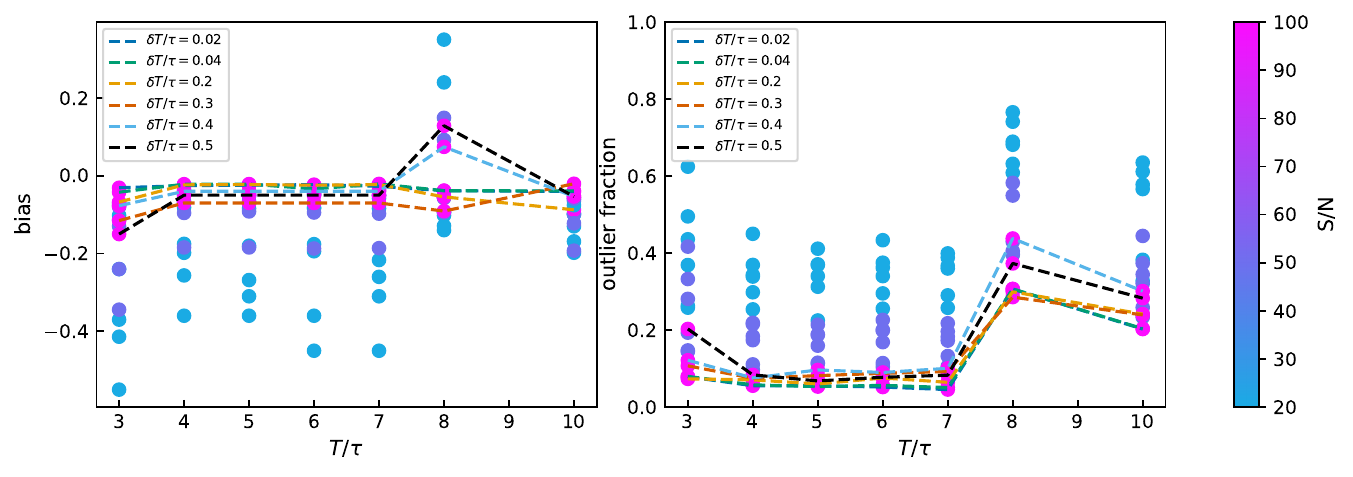}
    \caption{Bias and outlier fraction as a function of $T/\tau$ for the three simulation trials. \textit{Top to bottom}: Trial 1, trial 2, and trial 3. Dashed coloured lines connect points with the same $\delta T/\tau$ for visual guidance; only the S/N = 100 points are connected.}
    \label{fig:bias_trial}
\end{figure*}

In all set-ups, the recovered lag was compared to the input lag using
the bias, $b$, and the fraction of simulations classified as outliers
(defined as recoveries deviating by more than 50\% from the true lag).
Our key findings are as follows:

Trial 1 demonstrated the importance of an appropriately defined lag search range. 
Because seasonal gaps shorten the light curves relative to the nominal baseline, $T$, a search range of $[-0.3 T,0.5 T]$ enabled the ICCF to explore lags that could not be sampled at least twice within the data, a common requirement for reliable cross-correlation. 
This degraded the recovery around $T / \tau \sim 7$, where a gap falls towards the end of the simulated campaign (Fig.~\ref{fig:bias_trial} top panel). 
The baseline-based definition was originally chosen to reflect the sparse metadata on seasonal gaps in published campaigns. For all subsequent simulations, the search range is instead defined from the actual light-curve length.

Trials 2 and 3 show that the ICCF generally underestimates the delay
and that the S/N plays a critical role: at S/N~$< 20$ the outlier fraction
rises steeply (Fig.~\ref{fig:bias_trial}, middle panel). 
Smaller cadence ratios $\delta T / \tau$ favour the recovery at all baselines. 
For sufficiently long baselines we find no significant dependence on $T / \tau$ itself; short baselines combined with sparse sampling, however, strongly degrade the recovery. 
Applying a detrending step where the ICCF appears flat generally reduces the bias for long-baseline campaigns (Fig.~\ref{fig:bias_trial} bottom panel) but occasionally introduces artifacts for short baselines, leading us to exclude detrending from the main H$\beta$ simulation (Sect.~\ref{sec:sim_hbeta}).

\section{Application to the H$\beta$ $R$-$L$ relation}
\label{sec:rl_analysis}

\subsection{Expected lag from the reference $R$-$L$ relation}
\label{sec:expected_lag}
 
To assess whether a published lag could have been recovered under the
reported observational conditions, we first need an expected lag for
each source.
We computed the observer-frame expected lag from the source's continuum
luminosity using the reference relation,
\begin{equation}
  \label{eq:expected_lag}
  \log\!\left(\tau_{\rm obs}\right) = 1.5 + 0.5\,
  \log\!\left(\frac{L_{5100}}{10^{44}\,\mathrm{erg\,s^{-1}}}\right)
  + \log(1+z),
\end{equation}
which corresponds to a photoionisation slope of $\alpha = 0.5$ and
an offset approximately consistent with various calibrations
\citep[e.g.][]{Hu2025,Shen2024}.
The expected lag was used as the input lag for the corresponding
simulation.
We acknowledge that this step introduced a dependence on the assumed
reference relation, a caveat discussed further in
Sect.~\ref{sec:discussion}.

\subsection{Simulation of the H$\beta$ sample}
\label{sec:sim_hbeta}
 
For each of the 248 H$\beta$ sources for which sufficient observational
metadata are available, we ran 1000 DRW light-curve realisations with
the expected input lag, sampled them according to the campaign
parameters, applied the ICCF with the optimised settings
($t_{\rm unit} = 0.8\,\delta T$, $r = 0.8\,r_{\rm max}$), and recorded
the median retrieved lag $\tau_{\rm ret}$.
The variability amplitude for each source was estimated from
Eq.~\eqref{eq:sf} using its redshift, luminosity, and baseline.
Detrending was not applied; as noted above, it was found to be
unreliable over the broad range of observational parameters encountered
in the sample and did not significantly improve the ICCF performance.
Seasonal gaps were inserted as described in Sect.~\ref{sec:obs_sampling}.

The final simulation sample spans a redshift range of
$0 < z \leq 1.03$ (mean $z = 0.36$) and continuum
luminosities from $9.1\times10^{41}$ to
$4.4\times10^{45}\,\mathrm{erg\,s^{-1}}$, with a mean of
$3.2\times10^{44}\,\mathrm{erg\,s^{-1}}$.
The mean S/N of the sample is 56 (range 1-423).
More than half of the sources come from \citet{Shen2024} and
\citet{Bentz2013}.

\subsection{Flagging scheme}
\label{sec:flagging}
 
From each simulation, we recorded three lag values for every source:
the expected lag $\tau_{\rm exp}$ (computed from the $R$-$L$
relation), the reported lag $\tau_{\rm rep}$ in the literature (taken from the database),
and the simulation-retrieved lag $\tau_{\rm ret}$. We defined three dimensionless ratios of
\begin{equation}
  q_3 = \frac{\tau_{\rm exp}}{\tau_{\rm ret}},
  \qquad
  q_2 = \frac{\tau_{\rm rep}}{\tau_{\rm ret}},
  \qquad
  q_1 = \frac{\tau_{\rm exp}}{\tau_{\rm rep}}.
\end{equation}
 
A four-tier classification was applied exclusively. Specifically, a source was
assigned to the highest category whose condition it satisfies as follows: 
 
\begin{itemize}
  \item \textbf{Category~3} (highest risk): the ICCF fails to
        recover the expected lag, i.e.\
        $q_3 > 1.5$ or $q_3 < 0.5$.
        This indicates a direct inconsistency between the expected lag
        and our simulation, suggesting the observational campaign was
        unable to reliably constrain the lag.
  \item \textbf{Category~2}: $q_2 > 1.5$ or $q_2 < 0.5$.
        The simulation recovers a lag inconsistent with the reported
        value, suggesting a potential lag-recovery issue.
  \item \textbf{Category~1}: $q_1 > 1.5$ or $q_1 < 0.5$.
        The reported lag deviates significantly from the expected lag
        of the reference $R$-$L$ relation.
        This category is not linked to the simulation itself, but
        identifies sources that are strong outliers relative to the
        assumed relation.
  \item \textbf{Category~0}: No flag. The source is consistent with
        the assumed reference relation and the simulation indicates
        reliable lag recoverability.
\end{itemize}

The thresholds of 0.5 and 1.5 were intentionally set to be conservative, so that sources were
only flagged when there was a clear indication of inconsistency. 
The adopted quality thresholds are intentionally asymmetric with respect 
to the evaluated ratios. This choice reflects practical considerations
related to ICCF lag recovery and is discussed further in Sect.~\ref{sec:asymmetric}.
 
The resulting category distribution is shown in
Table~\ref{tab:categories}.
A total of 143 sources ($\simeq 58\%$) fall in category~0;
94 ($\simeq 38\%$) in category~2; and 11 ($\simeq 5\%$) in
category~3.
No source is assigned to category~1, indicating that no source in
the sample is an extreme outlier relative to the reference relation
beyond what is already captured by the simulation-based categories. A table containing the simulation results of each AGN can be viewed online (see Table~\ref{tab:sim_summary}).
 
\begin{table}[htbp]
  \caption{Distribution of the 248 H$\beta$ sources among the four
           flagging categories.}
  \label{tab:categories}
  \centering
  \begin{tabular}{lcc}
    \hline\hline
    Category & $N$ & Fraction (\%) \\
    \hline
    0 (no flag)         & 143 & 57.7 \\
    1 ($R$-$L$ outlier)  &   0 &  0.0 \\
    2 (sim.\ discrepancy) &  94 & 37.9 \\
    3 (ICCF failure)    &  11 &  4.4 \\
    \hline
    Total               & 248 & 100  \\
    \hline
  \end{tabular}
\end{table}

\subsection{Refitting the $R$-$L$ relation}
\label{sec:refitting}
 
\begin{figure*}
    \centering
    \includegraphics[width=0.50\linewidth]{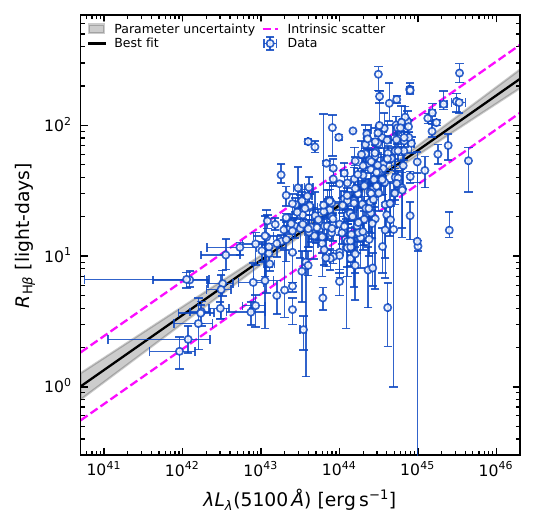}
    \includegraphics[width=0.49\linewidth]{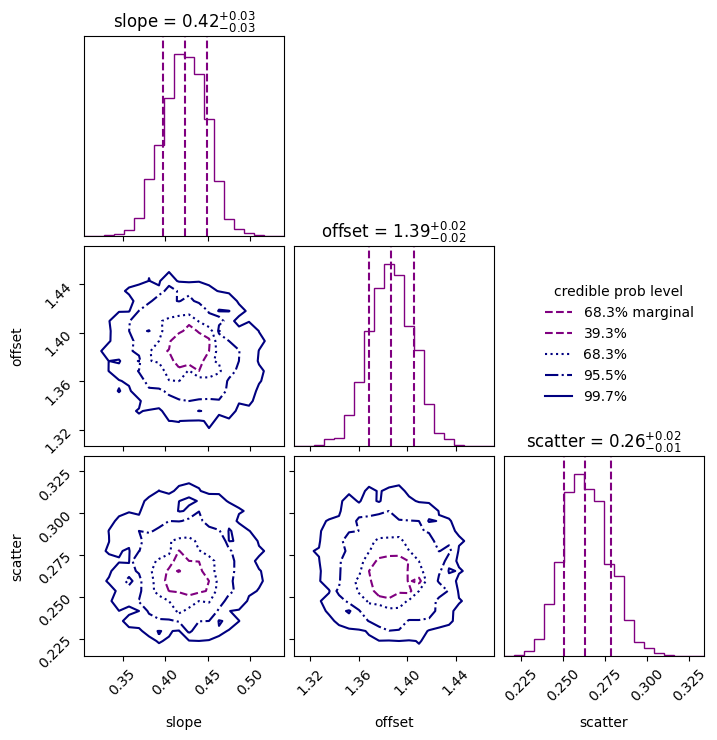}
    \caption{\textit{Left:} Best fit of the H$\beta$ $R$-$L$ relation for the full sample. The solid~black line shows the best-fit relation, the grey shaded area its uncertainty, and the dashed magenta lines the fitted intrinsic scatter. \textit{Right:} Posterior distributions and parameter correlations for the same fit.}
    \label{fig:fit_all}
\end{figure*}

\begin{figure}
    \centering
    \includegraphics[width=1\linewidth]{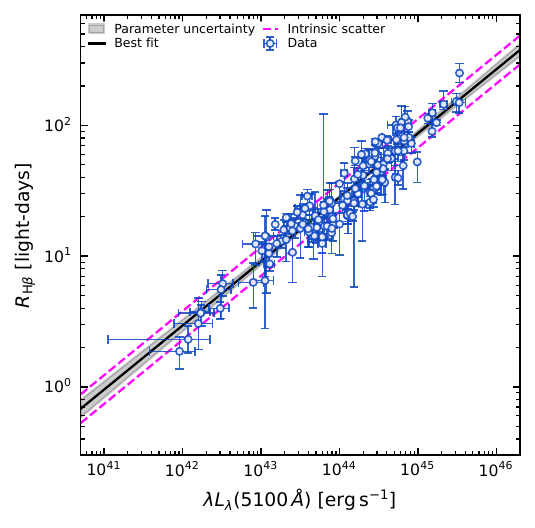}
    \caption{Best-fit of the H$\beta$ $R$-$L$ relation for the sample
    consisting of Cat.\ 0 assuming $\alpha_{\rm ref} = 0.5$ and thresholds 
    0.5/1.5. The solid~black lines shows the best linear fit, gray shaded areas 
    indicate the uncertainty of the fit, dashed magenta lines show fitted intrinsic 
    scatter.}
    \label{fig:fit_cat_0}
\end{figure}

We refitted the $R$-$L$ relation in the form of Eq.~\eqref{eq:rl} using the Bayesian nested-sampling
code UltraNest\footnote{\url{https://johannesbuchner.github.io/UltraNest/}}
\citep{Buchner2021} with uniform
priors $\alpha \in [0, 2]$, $\beta \in [-5, 5]$, and
$\sigma \in [0, 1]$.
Each data point was sampled 100\,000 times assuming Gaussian errors.
The sampler was run with a minimum of 30 live points and targeted a
minimum of 1000 effective posterior samples. We fitted four progressively cleaner subsamples:

\begin{enumerate}
  \item All 248 sources (categories 0-3; see Fig.~\ref{fig:fit_all}).
  \item Categories 0, 1, 2 (excluding category~3; $N = 237$).
  \item Categories 0 and 1 (excluding categories 2 and 3; $N = 143$).
  \item Category~0 only ($N = 143$; see Fig.~\ref{fig:fit_cat_0}).
\end{enumerate}
Because no source falls in category~1, samples 3 and 4 are identical
for this particular reference model.

Table~\ref{tab:rl_fits} summarises the best-fit parameters for the
reference slope of $\alpha_{\rm ref} = 0.5$.
Removing only the 11 category-3 sources has a minor effect on
the scatter ($\sigma$ decreases from 0.26 to 0.25~dex).
However, removing the additional 94 category-2 sources drives a
dramatic reduction to $\sigma = 0.11\pm0.01$~dex, with a corresponding
increase in slope from $\alpha = 0.42$ to $0.49\pm0.02$ (see Fig.~\ref{fig:fit_cat_0}). The fits
and their posteriors of all subsamples are shown in Appendix~\ref{app:fits}.

\begin{table}[htbp]
  \caption{Best-fit parameters of the H$\beta$ $R$-$L$ relation
           (Eq.~\ref{eq:rl}) for progressive subsamples.
          }
  \label{tab:rl_fits}
  \centering
  \begin{tabular}{lcccc}
    \hline\hline
    Sample & $N$ & $\alpha$ & $\beta$ & $\sigma$ \\
           &     &          &         & (dex)    \\
    \hline
    All         & 248 &
      $0.42\pm0.03$ & $1.39\pm0.02$ & $0.26^{+0.02}_{-0.01}$ \\
    Cat.\ 0+1+2 & 237 &
      $0.44^{+0.03}_{-0.02}$ & $1.40\pm0.02$ & $0.25\pm0.02$ \\
    Cat.\ 0+1   & 143 &
      $0.49\pm0.02$ & $1.45\pm0.01$ & $0.11\pm0.01$ \\
    Cat.\ 0     & 143 &
      $0.49\pm0.02$ & $1.45\pm0.01$ & $0.11\pm0.01$ \\
    \hline
  \end{tabular}
  \tablefoot{ The reference model assumes $\alpha_{\rm ref} = 0.5$.}
\end{table}

\subsubsection{Sensitivity to flagging thresholds}

To test the sensitivity to the flagging thresholds, we applied both
a stricter scheme with boundaries at 1.3/0.7 instead of 1.5/0.5,
and a looser scheme with boundaries at 1.7/0.3.
The resulting category distribution is shown in 
Table~\ref{tab:categories_strict_loose} for the stricter and looser
scheme.

\begin{table}[htbp]
  \caption{Distribution of the 248 H$\beta$ sources among the four
           categories using the stricter flagging thresholds (Col. 2-3)
           and looser flagging thresholds (Col. 4-5).}
  \label{tab:categories_strict_loose}
  \centering
  \resizebox{0.5\textwidth}{!}{
  \begin{tabular}{lcccc}
    \hline\hline
     & \multicolumn{2}{c}{strict thresholds} & \multicolumn{2}{c}{loose thresholds} \\
    \cmidrule(r){2-3}\cmidrule(l){4-5}

     Category & $N$ & Fraction (\%) & $N$ & Fraction (\%) \\

    \hline
    0 (no flag)         & 87 & 35.1 & 197 & 79.4 \\
    1 ($R$-$L$ outlier)  &   3 &  1.2 &   2 &  0.9\\
    2 (sim.\ discrepancy) &  145 & 58.5 &  39 & 15.7 \\
    3 (ICCF failure)    &  13 &  5.2 &  10 &  4.0 \\
    \hline
    Total               & 248 & 100  & 248 & 100\\
    \hline
  \end{tabular}}
\end{table}

The most restrictive sample yields $\sigma = 0.05\pm0.01$ and 
$\alpha = 0.49\pm0.01$ using the stricter threshold, and 
$\sigma = 0.16\pm0.01$ and $\alpha = 0.48\pm0.02$ using the 
looser threshold. Table~\ref{tab:rl_fits_strict_loose}
summarises these results.

\begin{table*}[htbp]
  \caption{Best-fit parameters of the H$\beta$ $R$-$L$ relation 
           (Eq.~\ref{eq:rl}) for the strict flagging scheme 
           (thresholds 1.3/0.7) and loose flagging scheme
           (thresholds 1.7/0.3).}
  \label{tab:rl_fits_strict_loose}
  \centering
  \begin{tabular}{lcccccccc}
    \hline\hline
         & \multicolumn{4}{c}{strict thresholds} & \multicolumn{4}{c}{loose thresholds} \\
    \cmidrule(r){2-5}\cmidrule(l){6-9}
    Sample & $N$ & $\alpha$ & $\beta$ & $\sigma$& $N$ & $\alpha$ & $\beta$ & $\sigma$ \\
           &     &          &         & (dex)&     &          &         & (dex)       \\
    \hline
    All & 248 & $0.42\pm0.03$ & $1.39\pm0.02$ & $0.26^{+0.02}_{-0.01}$ 
    & 248 & $0.42\pm0.03$ & $1.39\pm0.02$ & $0.26^{+0.02}_{-0.01}$ \\
    Cat.\ 0+1+2 & 235 & $0.44\pm0.03$ & $1.40\pm0.02$ & $0.26\pm0.01$
    & 238 & $0.45\pm0.03$ & $1.40\pm0.02$ & $0.25\pm0.01$\\
    Cat.\ 0+1 & 90 & $0.49\pm0.01$ & $1.49\pm0.01$ & $0.05\pm0.01$
    & 199 & $0.48\pm0.02$ & $1.39\pm0.01$ & $0.17\pm0.01$\\
    Cat.\ 0 & 87 & $0.49\pm0.01$ & $1.48\pm0.01$ & $0.05\pm0.01$ 
    & 197 & $0.48\pm0.02$ & $1.39\pm0.01$ & $0.16\pm0.01$\\
    \hline
  \end{tabular}
  \tablefoot{Reference model: $\alpha_{\rm ref} = 0.5$.}
\end{table*}

The fitted slope is consistent within the error bars regardless of the adopted thresholds, whereas the intrinsic scatter depends strongly on the selection. 
We therefore retain the initial thresholds of 0.5/1.5 for the further analysis: this selection balances the strict and loose alternatives, accounting for genuine recovery inconsistencies without artificially reducing the scatter towards the reference model.

\subsubsection{Model bias}

To investigate whether our approach introduced a bias towards our assumed
reference model (see Eq.~\ref{eq:expected_lag}), we re-simulated the sample
assuming a slope $\alpha_{\rm ref} = \frac{1}{3}$ and $\alpha_{\rm ref} = 
0.7$, commonly reported as the lower and upper bounds in the literature \citep[]{Kaspi2000, Shen2013}.
The distribution in the categories is shown in 
Table~\ref{tab:categories_03_07} for both $\alpha_{\rm ref}=\frac{1}{3}$ and 
$\alpha_{\rm ref}=0.7$.

\begin{table}[htbp]
  \caption{Distribution of the 248 H$\beta$ sources among the four
           flagging categories assuming a reference slope of $\alpha_{\rm ref}=\frac{1}{3}$ (Col. 2-3) and assuming a reference slope
           of $\alpha_{\rm ref}=0.7$ (Col. 4-5).}
  \label{tab:categories_03_07}
  \centering
  \resizebox{0.5\textwidth}{!}{
  \begin{tabular}{lcccc}
    \hline\hline
         & \multicolumn{2}{c}{$\alpha_{\rm ref} = \frac{1}{3}$} 
         & \multicolumn{2}{c}{$\alpha_{\rm ref} = 0.7$} \\
    \cmidrule(r){2-3}\cmidrule(l){4-5}
    Category & $N$ & Fraction (\%) & $N$ & Fraction (\%)  \\
    \hline
    0 (no flag)         & 136 & 54.8 & 119 & 48.0 \\
    1 ($R$-$L$ outlier)  &   5 &  2.1 &   7 &  3.0  \\
    2 (sim.\ discrepancy) &  97 & 39.1 &  110 & 44.2\\
    3 (ICCF failure)    &  10 &  4.0 &  12 &  4.8\\
    \hline
    Total               & 248 & 100 & 248 & 100 \\
    \hline
  \end{tabular}}
\end{table}

We repeated the fitting procedure, progressively excluding flagged objects, and summarise the results in Table~\ref{tab:rl_fits_03_07}.

\begin{table*}[htbp]
  \caption{Best-fit parameters of the H$\beta$ $R$-$L$ relation 
           (Eq.~\ref{eq:rl}) for the reference model with 
           $\alpha_{\rm ref} = \frac{1}{3}$ (Cols. 2-5) and
           $\alpha_{\rm ref} = 0.7$ (Cols. 6-9).}
  \label{tab:rl_fits_03_07}
  \centering
  \begin{tabular}{lcccccccc}
    \hline\hline
         & \multicolumn{4}{c}{$\alpha_{\rm ref} = \frac{1}{3}$} & \multicolumn{4}{c}{$\alpha_{\rm ref} = 0.7$} \\
    \cmidrule(r){2-5}\cmidrule(l){6-9}
    Sample & $N$ & $\alpha$ & $\beta$ & $\sigma$& $N$ & $\alpha$ & $\beta$ & $\sigma$ \\
           &     &          &         & (dex)&     &          &         & (dex)       \\
    \hline
    All & 248 & $0.42\pm0.03$ & $1.39\pm0.02$ & $0.26^{+0.02}_{-0.01}$ 
    & 248 & $0.42\pm0.03$ & $1.39\pm0.02$ & $0.26^{+0.02}_{-0.01}$ \\
    Cat.\ 0+1+2 & 238 & $0.45\pm0.03$ & $1.40\pm0.02$ & $0.25\pm0.01$
    & 236 & $0.44\pm0.03$ & $1.40\pm0.02$ & $0.26\pm0.01$\\
   Cat.\ 0+1 & 141 & $0.37\pm0.02$ & $1.43\pm0.01$ & $0.10\pm0.01$
    & 126 & $0.61^{+0.02}_{-0.03}$ & $1.43\pm0.01$ & $0.12\pm0.01$\\
    Cat.\ 0 & 136 & $0.37\pm0.02$ & $1.43\pm0.01$ & $0.10\pm0.01$ 
    & 119 & $0.62\pm0.02$ & $1.44\pm0.01$ & $0.11\pm0.01$\\
    \hline
  \end{tabular}
\end{table*}

The most restrictive samples show a bias towards the reference slope
in both cases. The recovered slope is $\alpha =0.37\pm0.02$ for $\alpha_{\rm ref} = \frac{1}{3}$ and $\alpha=0.62\pm0.02$ for $\alpha_{\rm ref} = 0.7$.
The latter case shows that, although a bias towards the reference model exists, the current H$\beta~R$-$L$ sample favours a shallower slope.
Conversely, with the shallower reference model the fitted slope lies above the reference value, indicating (albeit less significantly) a preference for a steeper slope.
Independently of the reference slope, the fitted intrinsic scatter agrees across all cases at $\sigma \approx 0.11$~dex.
We also find that the category-3 assignments largely overlap: most category-3 sources are flagged as such independent of the reference model or flagging thresholds. 
This strengthens the conclusion that, for these objects and observing conditions, the ICCF cannot deliver a reliable lag.

We note that for the steeper reference slope a mild anti-correlation
between the fitted slope and offset appears in the posterior, most
likely a fitting degeneracy driven by the reduced luminosity range
of the selected sample.

\subsubsection{Correcting for model bias}
\label{sec:model-bias-correction}
To test whether the reference-model bias can be compensated for in the refitting procedure, we constructed new samples by intersecting the classifications obtained under the different reference models:
\begin{enumerate}
    \item All sources where the different reference models jointly
          categorise the source in category 0.
    \item Exclusion of sources with flag 3 assigned by at least one reference
          model, and both joint categorisation in category 2 and category 1.
\end{enumerate}

\noindent
We also implemented a second approach to reduce the reference model bias
by counting the number of times, $N_{\rm good}$, that a source was assigned to category 0 by the different reference models. Depending on $N_{\rm good}$ we define additional subsamples:
\begin{enumerate}
    \setcounter{enumi}{2}
    \item Sources with $N_{\rm good}\geq2$
    \item Sources with $N_{\rm good}\geq1$
\end{enumerate}

We note that a sample requiring $N_{\rm good} = 3$ would be identical to sample 1 and was therefore not considered. In brief, sample 1 has poor luminosity coverage, inherited from the sparse category-0 population under $\alpha_{\rm ref} = 0.7$; sample 2 retains sources for which the reference models disagree on a simulation discrepancy, while excluding every source flagged as an ICCF failure by at least one model; sample 3 contains sources deemed low risk by the majority of reference models; and sample 4 contains all sources deemed low risk at least once, including some flagged as discrepant by two models.
We fitted the subsamples with the previously adopted
routine and show the results in Table~\ref{tab:rl_joint_fit}
and Fig.~\ref{fig:intersect_matrix}.

\begin{table}[htbp]
  \caption{Best-fit parameters of the H$\beta$ $R$-$L$ relation
           (Eq.~\ref{eq:rl}) for the new subsamples correcting
           for model bias.}
  \label{tab:rl_joint_fit}
  \centering
  \begin{tabular}{lcccc}
    \hline\hline
    Sample & $N$ & $\alpha$ & $\beta$ & $\sigma$ \\
           &     &          &         & (dex)    \\
    \hline
    1         & 85 &
      $0.51\pm0.03$ & $1.42\pm0.01$ & $0.08\pm0.01$ \\
    2 & 179 &
      $0.47\pm0.02$ & $1.45\pm0.01$ & $0.15\pm0.01$ \\
    3 & 136 &
      $0.48\pm0.02$ & $1.45\pm0.01$ & $0.11\pm0.01$ \\
    4   & 177 &
      $0.47\pm0.02$ & $1.45\pm0.01$ & $0.14\pm0.01$ \\
    \hline
  \end{tabular}
\end{table}

With both approaches, the range of recovered slopes narrows from $0.37$-$0.62$ to $0.47$-$0.51$, reducing the model dependence of the classification by a factor of $\sim$6. By judging each source jointly across the reference models, only sources whose classification is robustly model-independent are excluded.

We conclude that the inferred slope becomes substantially less 
sensitive to the classification procedure when only objects 
with consistent classifications across multiple reference 
relations are retained. While the original approach yielded 
slopes ranging from 0.37 to 0.62, the consensus-based selections 
produce values between 0.47 and 0.51, indicating that much of the 
model dependence originates from objects whose classification is 
itself sensitive to the assumed reference relation.

\section{Discussion}
\label{sec:discussion}
\subsection{Asymmetric flag thresholds}
\label{sec:asymmetric}

A possible limitation of the adopted quality thresholds is that the
ratio criterion is not mathematically symmetric: for
$q=\tau_{\rm exp}/\tau_{\rm ret}$, the limits $0.5<q<1.5$ do not permit
the same fractional deviation in both directions, which would instead
require $2/3<q<3/2$. We retain the asymmetric thresholds deliberately.
Since the ICCF more frequently underestimates than overestimates the
true lag (Sect.~\ref{sec:sim_parameterstudy}), symmetric limits would
disproportionately reject moderate underestimates — some of which are
flagged only because of the adopted reference offset — while leaving
the acceptance of overestimates essentially unchanged. The lower bound
of $0.5$ is thus a conservative allowance for this recovery bias.

We verified this by adopting the symmetric lower threshold of $2/3$.
Although only $\sim$15\% of measurements change category, the recovered
normalisation moves systematically closer to the assumed reference
relation and the inferred scatter decreases slightly, confirming that
the stricter bound preferentially removes moderate underestimates and
increases the model dependence of the result — which we regard as
undesirable given the systematic uncertainty of the reference relation
itself.

A similar fraction ($\sim$15\%) changes category if the inverse ratios
(e.g.\ $\tau_{\rm ret}/\tau_{\rm exp}$) are used instead, with roughly
half of the affected sources migrating from category~0 into category~1.
Since category~1 reflects a disagreement with the reference relation
alone, this definition would amplify sensitivity to the assumed model,
rather than to the quality of the lag recovery; therefore, we retained
the adopted ratio definitions.

\subsection{Constraining the $R$-$L$ relation}
Our initial analysis shows that the fitted
$R$-$L$ relation 
is strongly dependent on the preset
reference relation. In Sect.~\ref{sec:model-bias-correction},
we define new subsamples to correct for the dependency
by accounting for classification sensitivity and consensus
between the different reference models. The reduced range of
slope recovery for the subsamples from 0.37–0.62 to 0.47–0.51 after correcting for the model bias shows that
these samples are significantly less affected by the 
employed reference models.

We note that sample 1 is sensitive to the flagging thresholds. 
A first indication is its low intrinsic scatter, similar to that obtained with the stricter thresholds. 
Moreover, in the limit of very strict thresholds this sample retains only sources simultaneously consistent with all three reference relations, so its fitted slope necessarily tends towards the average of the reference slopes — and the fitted parameters indeed lie close to that average. 
This sample should therefore be interpreted with caution.

Constraining the $R$-$L$ relation requires balancing the correction of the model bias against reasonable quality cuts. 
Samples 2 and 4 primarily reduce the model bias but likely retain lower-quality measurements; we therefore adopt sample 3 as the final calibration sample. 
This choice is not critical: slope and offset are consistent across the subsamples and only the inferred intrinsic scatter depends on it. 
The scatter of sample 3 is similar to that of the individual reference-slope fits, in line with our conclusion that the adopted thresholds balance overly strict and overly loose cuts.

A comparison of this work's H$\beta~R$-$L$ calibration with
calibrations from the literature can be found in 
Fig.~\ref{fig:lit_comp_scatter}. 
Our calibration shows the
smallest uncertainties in the recovered slope and is located
in the middle of the selection. The slope of $\alpha=0.48\pm0.02$
is consistent with the photoionisation expectation.

\subsection{Reduction of intrinsic scatter}
\label{sec:scatter_discussion}
 
\begin{figure*}
    \centering
    \includegraphics[width=0.49\linewidth]{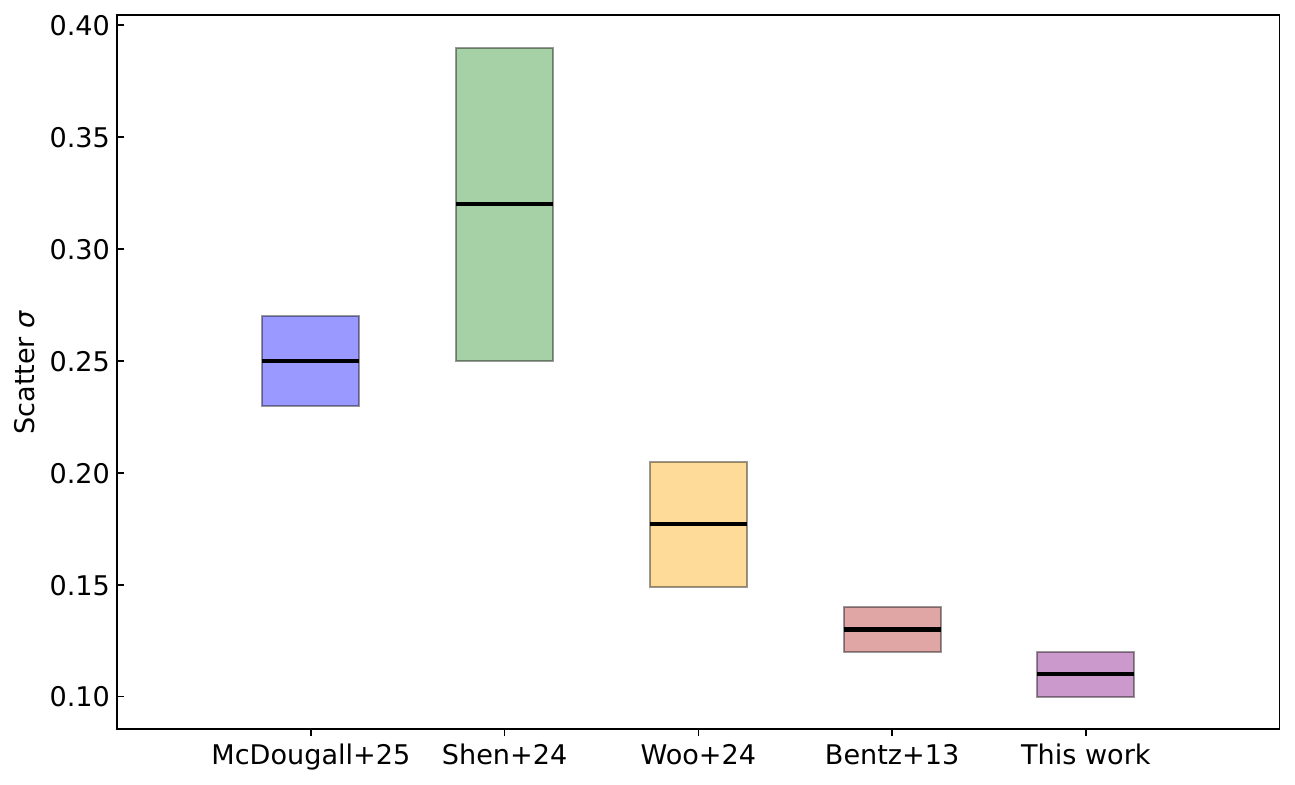}
    \includegraphics[width=0.49\linewidth]{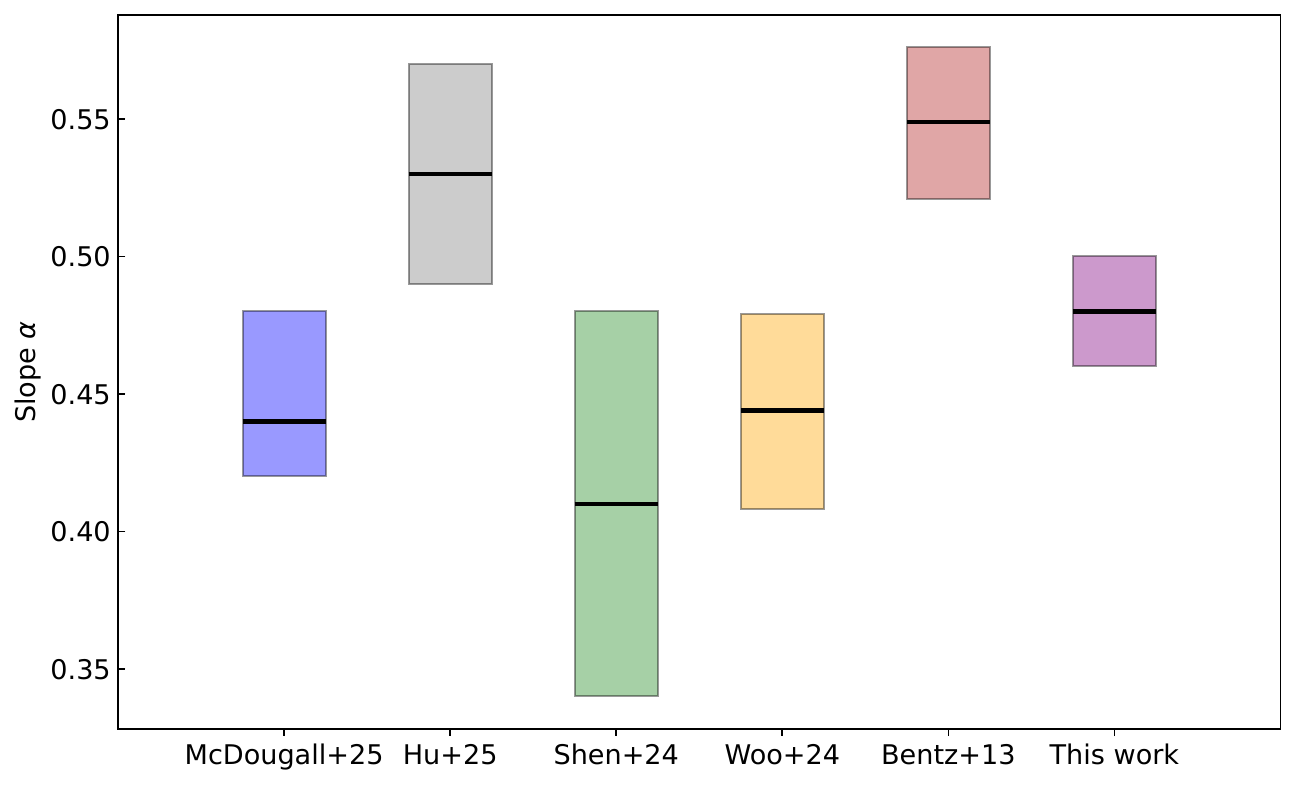}
    \caption{Comparison of the intrinsic scatter (\textit{left}) and slope (\textit{right}) of this work's
    H$\beta$ $R$-$L$ calibration (sample 3) with a selection from the literature. Black lines mark the fitted values; coloured boxes indicate the fitting uncertainties.}
    \label{fig:lit_comp_scatter}
\end{figure*}

The most robust result of our analysis is the reduction of the
inferred intrinsic scatter from $\sigma \approx 0.26$~dex for the
full sample to $\approx 0.11$~dex for the final, model-bias-corrected sample. 
Among the literature calibrations (Fig.~\ref{fig:lit_comp_scatter}), 
ours shows the smallest intrinsic scatter, suggesting that previous estimates were inflated by unreliable measurements.
Although a residual model dependence cannot be fully excluded, the same scatter is recovered for multiple cleaned subsamples, demonstrating the robustness of the result.
The intrinsic scatter is insensitive to the employed reference model and depends primarily on the flagging thresholds, for which our adopted values balance stricter and looser selections.

This large reduction has a clear physical interpretation: a significant
fraction of the observed scatter in the H$\beta$ $R$-$L$ relation
does not reflect genuine diversity in BLR sizes among AGNs of equal
luminosity, but instead arises from limitations in lag recovery.
 
Our simulations show that the ICCF systematically underestimates lags,
especially at low S/N.
Category-2 and category-3 sources disproportionately lie below the
$R$-$L$ relation at the high-luminosity end, consistent with an
underestimation of their (typically longer) lags.
Consequently, the reduced scatter likely arises from accounting 
for the systematic underestimation of lags in risky measurements, 
rather than from the reduction of the sample itself. 
Alternatively, this could indicate that a linear $R$-$L$ relation may
not be correct, but that analysis is beyond the scope of this paper.

These findings support the conclusion that improving the quality
control of RM samples is essential before the scatter of the $R$-$L$
relation can be used to draw conclusions about the physical diversity
of the BLR.

\subsection{Implications for single-epoch mass estimators}
\label{sec:se_mass}

SE virial mass estimators remain the primary method
of estimating black hole masses at high redshift, where dedicated
RM campaigns are generally unavailable.
Two approaches are commonly used to calibrate them: fitting RM data
directly to a function of luminosity and line width taking the form
$\log M_{\rm BH} = a + b\log L + c\log W$ or substituting an
empirical $R$-$L$ relation into the virial mass equation
(Eq.~\ref{eq:virial}).
The first approach was adopted, for example, by \citet{Shen2024}. It requires the underlying $R$-$L$ relation to remain tightly
constrained when additional free parameters are introduced; however, this is a condition that is not guaranteed and that can reintroduce scatter.
We therefore adopted the second approach, which allowed us to propogate our
bias-corrected $R$-$L$ calibration directly into the mass estimate.

Using the best-fitting parameters of the consensus bias-corrected
sample 3 of Sect.~\ref{sec:refitting}, we obtained the H$\beta$ SE mass
estimator,
\begin{equation}
  \label{eq:hb_se}
  \frac{M_{\rm BH}}{M_\odot} =
  5.469\, f\,
  \left(\frac{L_{5100}}{10^{44}\,\mathrm{erg\,s^{-1}}}\right)^{0.48}
  \left(\frac{W}{\mathrm{km\,s^{-1}}}\right)^2 ,
\end{equation}
where $W$ is the H$\beta$ line width.
As discussed, we adopt this sample because it spans a 
broader luminosity range than the strictest selection and is 
less affected by classification boundary effects, while retaining
substantially reduced model bias relative to the single 
reference-model fits.

We deliberately refrain from prescribing a virial factor $f$ or a
particular line-width measure.
The virial factor is an individual property of each AGN that is not
well predicted by population averages
\citep[e.g.][]{Onken2004,Francisco2014}, and attempts to link it to
observable quantities have so far not yielded tight correlations
\citep{Grier2017b,Williams2018,Villafana2026}.
The inferred mass also depends on whether the FWHM or the line
dispersion $\sigma_{\rm line}$ is adopted, with the two measures
introducing systematic offsets relative to one another
\citep{Peterson2004,Shen2013}.
We therefore recommend interpreting Eq.~(\ref{eq:hb_se}) primarily as
a virial-product estimator: any internally consistent line-width
metric may be used, while source-specific virial corrections can be
applied separately when independent constraints justify them.
Masses derived with different line-width indicators should not be
compared directly until a well-constrained conversion between FWHM
and $\sigma_{\rm line}$ becomes available.

\subsection{Implications for high-redshift black hole masses}
\label{sec:highz}
 
The motivation for this work partly originates from the large and
sometimes unexpected SMBH masses inferred at high redshift (see \citealt{Inayoshi2020} for a review).
Our findings suggest that some SE masses calibrated on the current $R$-$L$ relation
carry inflated uncertainties inherited from low-quality RM measurements.
In addition, calibrations based on smaller samples with narrower luminosity coverage may be biased by selection effects and by the systematic lag underestimation of the ICCF. 
Our broad compilation and quality selection mitigate these effects while accounting for model bias, allowing us to revise the black hole masses with the newly calibrated estimator (Eq.~\ref{eq:hb_se}). 
To ensure a fair comparison, we calculate
the SE mass estimators based on literature $R$-$L$ calibrations
using the same assumptions about velocity indicators and virial 
coefficients.

\begin{figure}
    \centering
    \includegraphics[width=1\linewidth]{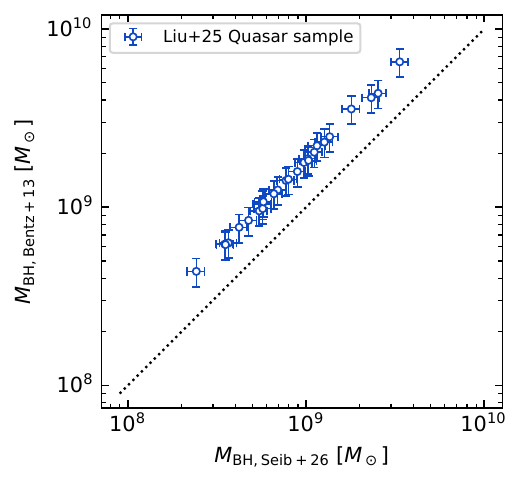}
    \caption{Comparison between this work's
    single epoch mass estimator and the single
    epoch mass estimator using \citet{Bentz2013}
    relation. The high $z$ quasar sample is from
    \citet{Liu2025}.}
    \label{fig:mass_comparison_bentz}
\end{figure}

Using the $R$-$L$ calibration of \citet{Bentz2013} yields 
the SE mass estimator 
$\frac{M_{\rm BH}}{M_\odot}
= 7.064\,f\,
\left(\frac{L_{5100}}{10^{44}\,\mathrm{erg\,s^{-1}}}\right)^{0.549}
\left(\frac{W}{\mathrm{km\,s^{-1}}}\right)^2$.
To assess the impact of the revised calibration, we can apply 
both estimators to the high $z$ quasar sample of \citet{Liu2025} ($z\sim5-6$) and compare 
the inferred black hole masses. Further comparisons with other BH mass
estimates are discussed in Appendix~\ref{app:mass_comp}.

The new calibration systematically yields lower mass estimates, 
with a typical reduction of approximately a factor of two relative 
to the \citet{Bentz2013} calibration. While the smaller slope 
leads to an increasingly larger reduction at high luminosities, 
the dominant contribution originates from the lower offset $\beta$ of 
the revised $R$-$L$ relation. The resulting mass difference exceeds 
the intrinsic scatter of both calibrations ($\sim0.13$ and 
$\sim 0.11$ dex), indicating that the mass difference cannot be attributed to the calibration uncertainties.

Nevertheless, the revised calibration alone is unlikely to resolve 
the tension posed by the highest inferred black hole masses. 
Current theoretical models would require a reduction of roughly 
an order of magnitude, substantially larger than the factor of 
$\sim2$ obtained with the updated calibration. In addition, the 
high observed bolometric luminosities of $z>7$ quasars would mean 
that the implied radiative efficiencies need to be $\gtrsim20\%$ 
if the black hole masses were smaller by a factor of two. Measurements 
of the radiative efficiency from ionised gas in their surroundings 
instead prefer values $<1\%$ \citep{Davies2019}.

Finally, the estimator and the $R$-$L$ relation should be tested on an AGN population at $z > 1$ to investigate possible population bias. We encourage dedicated RM campaigns in this regime, especially at high luminosities.

\subsection{Limitations}
\label{sec:limitations}
 
Our framework rests on several simplifying assumptions.
We model the cadence as uniform and the seasonal gap as a regular
$\sim$180-day window, whereas real campaigns experience irregular gaps due
to weather, scheduling, and technical issues.
The DRW variability amplitude is estimated from the global structure
function scaling \citep{Morganson2014}, rather than from the measured
light curves and the required reference $R$-$L$ model is assumed to be linear, without any turnover at a certain luminosity.
The ICCF parameters are fixed to a single optimised set; in practice,
analysts sometimes tune the lag search range, centroid threshold, and
interpolation settings for individual sources.
Finally, the database metadata are sometimes incomplete: cadences
are often not reported, and S/N estimates are heterogeneous.
Missing values are filled with default assumptions that may not
accurately represent the true observing conditions.
Despite these limitations, the framework can simulate 96\% of
the H$\beta$ sample and produces physically reasonable results.
It should be regarded as a quality filter for large samples, rather than as a definitive assessment of individual measurements.

The final H$\beta$ $R$-$L$ calibration was corrected for the strong reference-model bias found in the individual fits. 
By construction, the simulation always relies on a reference relation and the corrected sample can therefore never be fully independent of it; the consensus approach reduces this dependence to a minimal level.

\section{Conclusions}
\label{sec:conclusions}

We present the \texttt{pyRMTools} Python package, comprised of a database of published RM measurements, a simulation-based lag-recovery consistency framework, and the campaign-planning tool \texttt{scout}, together with a recalibrated H$\beta$ $R$-$L$ relation and SE mass estimator. The main findings and deliverables of this work are as follows:

\begin{enumerate}
 
\item \texttt{pyRMTools} incorporates a database of RM measurements covering
  $\sim$1200 AGNs from 32 campaigns spanning 1994-2026.
  The database stores rest-frame lags, continuum luminosities,
  line widths, virial products, black hole masses, and observational
  metadata in a hierarchical document structure accessible via Python.
  Utility functions for common retrieval tasks are provided.
  The database is publicly available and has been designed to be extendable with
  new campaigns.
 
\item A simulation-based consistency framework for assessing
  the recoverability of published H$\beta$ lags: the DRW light curves are sampled according to each campaign's
  observational parameters and analysed with the ICCF. The retrieved
  lag is compared to the expected and reported values using a
  four-tier flagging scheme. The key results in this regard are:
  \begin{itemize}
    \item The ICCF has a negligible bias ($< 0.35\%$) under ideal
          conditions, but systematically underestimates lags, especially
          when S/N $\lesssim 20$.
    \item In the H$\beta$ $R$-$L$ sample $\approx 40\%$ of sources
          (category~2) show a discrepancy between the reported and
          simulation-retrieved lags; $\approx 5\%$ (category~3) show
          a direct inconsistency between the expected and retrieved
          lags.
    \item Refitting the $R$-$L$ relation after excluding flagged
          sources and correcting for model bias 
          reduces the intrinsic scatter from
          $\sigma = 0.26^{+0.02}_{-0.01}$~dex to
          $\sigma = 0.11\pm0.01$~dex, with a slope of
          $\alpha = 0.48\pm0.02$ consistent with the photoionisation
          expectation.
    \item The scatter reduction is robust across different reference
          models and flagging thresholds; the slope depends on the
          assumed reference model and fully correcting for that 
          dependency is likely impossible.
          The resulting slope should not be interpreted as a
          direct measurement. It should instead be regarded as a model-bias-corrected calibration; the consensus construction bounds the dependence, yielding
          $\alpha = 0.48 \pm 0.02$, consistent with the photoionisation expectation.
  \end{itemize}
  \item \texttt{scout}, a planning tool for future RM campaigns built on the consistency framework. Given the planned observing parameters, it provides a first assessment of whether a reliable lag recovery can be expected, optimising the use of telescope time.
     
  \item A newly calibrated SE black hole mass estimator based on the 5100\AA~continuum luminosity and the H$\beta$
        line width. This would yield mass estimates that are typically a factor of $\sim$2 lower at high redshift than the calibration of \citet{Bentz2013}.
 
\end{enumerate}
 
Our results indicate that a significant fraction of the observed
scatter in the H$\beta$ $R$-$L$ relation arises from observational
limitations and lag-recovery biases, rather than any intrinsic AGN
diversity. A simulation-based quality control tool offers a practical way to
identify unreliable measurements without reanalysing the original light
curves. We encourage future RM campaigns to adopt a consistent method of reporting of the
baselines, cadences, and S/N values to enable this type of
retrospective quality assessment.

\section*{Data availability}
The full simulation results table (Table~\ref{tab:sim_summary}), together
with the analogous tables for all tested reference slopes and flagging
thresholds, are available in electronic form at the CDS via \url{http://cdsweb.u-strasbg.fr/cgi-bin/qcat?J/A+A/...}.
The reverberation mapping database is distributed as a JSON archive
together with the open-source Python package \texttt{pyRMTools}, which
also contains the simulation framework and the campaign-planning tool
\texttt{scout}. The package, the database, installation instructions,
and tutorials reproducing the analyses of this paper are available at
\url{https://github.com/Juri-W-S/pyRMTools.git}.
Full-resolution versions of the fit and posterior matrices
(Figs.~\ref{fig:fit_matrix}-\ref{fig:intersect_matrix}) will be provided
in the same repository.

\begin{acknowledgements}
F. Pozo Nu\~nez gratefully acknowledges the generous and invaluable support of the Klaus Tschira Foundation.
F. Pozo Nu\~nez acknowledge funding from the European Research Council (ERC) under the European Union's Horizon 2020 research and innovation program (grant agreement No 951549). 
SEIB is supported by the Deutsche Forschungsgemeinschaft (DFG) under Emmy Noether grant number BO 5771/1-1a.
We thank the anonymous referee for a constructive report that helped improve the clarity of this paper.
This research has made use of the NASA/IPAC Extragalactic Database (NED) which is operated by the Jet Propulsion Laboratory, California Institute of Technology, under contract with the National Aeronautics and Space Administration.
This research has made use of the SIMBAD database, operated at CDS, Strasbourg, France.
This research has made use of \texttt{Astropy}, a community-developed core Python package for astronomy \citep{astropy}, as well as
\texttt{Matplotlib} \citep{matplotlib}, \texttt{Numpy} \citep{numpy}, \texttt{Pandas} \citep{pandas} and \texttt{Scipy}
\citep{scipy}.
      
\end{acknowledgements}

\bibliographystyle{aa}
\bibliography{RM-bibliography}

\begin{appendix}

\section{pyRMTools structure and API overview}
\label{app:pyRMTools}

We give an overview of the structure and API developed for
the \texttt{pyRMTools} package. Future updates and step-by-step tutorials demonstrating the package's capabilities are available in the GitHub repository.

\subsection{Database class}

The Database class provides the entry point to the complete reverberation 
mapping database. Besides importing the data, it serves as the central interface 
for querying AGNs and measurements. Whether the data are loaded from the distributed 
JSON archive or from a MongoDB back-end, the user interacts with the database 
through the same API. The JSON back-end can be loaded using

\begin{verbatim}
import pyRMTools as qrm
db = qrm.Database.from_json()
\end{verbatim}

and the MongoDB back-end using

\begin{verbatim}
import pyRMTools as qrm
db = qrm.Database.from_mongodb(url, 
            database, collection)
\end{verbatim}

For the latter, the database must first be hosted in a local MongoDB instance; a tutorial is provided in the GitHub repository. 
This option is recommended only for users who wish to modify the back-end structure and are experienced with databases.

\subsubsection{Querying the database}

Measurements can either be queried globally from the complete database 
or per object. Global queries are useful when constructing samples 
for statistical analyses.

\begin{verbatim}
lag_collection = db.lag('H_beta')
\end{verbatim}

This returns a \texttt{LagCollection} containing every H$\beta$ lag 
measurement available in the database, independent of the parent AGN. 
Each measurement still contains a reference to its originating AGN through 
the \texttt{parent} attribute, allowing for object-specific information to 
be accessed when required. To access the individual measurements
from the collection, it is sufficient to iterate through and
call the desired properties.

\begin{verbatim}
for lag in lag_collection:
    print(lag.name, lag.value)
\end{verbatim}  

Equivalent methods are available for line widths, luminosities, 
black hole masses and virial products.

Alternatively, the database can be iterated over object-by-object
when analyses require access to several observables of the same AGN.

\begin{verbatim}
lags = []
luminosities = []

for agn in db:

   lag_measurements = agn.lag("H_beta")
   luminosity_measurements = agn.luminosity(5100)

   if lag_measurements and luminosity_measurements:
        lags.extend(
            lag_measurements.measurements)
        luminosities.extend(
            luminosity_measurements.measurements)
\end{verbatim}

This approach is particularly convenient when combining different 
measurements belonging to the same object, such as lags, luminosities, 
line widths, and black hole masses.

\subsection{AGN class}

Each AGN is represented by an \texttt{AGN} object that provides
access to all measurements available for this source. The class 
itself intentionally contains only object-level information, such as 
common identifiers, coordinates, and redshifts. Individual
observables are organised into dedicated \texttt{MeasurementCollection}
classes, which allows the AGN interface to remain consistent if
new measurement types are added.

\subsubsection{Measurements}
\label{app:measurements}
Every physical quantity in the database inherits the common \texttt{Measurement}
base class. Therefore, all measurements expose a common interface independent
of whether they represent a lag, luminosity or line-width. Every measurement 
contains
\begin{itemize}
    \item \texttt{measurement.entry}: entire entry of the json file.
    \item \texttt{measurement.value}: the value of the measurement.
    \item \texttt{measurement.unit}: the unit the measurement is taken in.
    \item \texttt{measurement.source}: the link to the publication reporting the measurement.
    \item \texttt{measurement.problematic}: the boolean flag implemented by us.
    \item \texttt{measurement.parent}: reference to the parent AGN object.
\end{itemize}

and the following properties depending on availability:

\begin{itemize}
    \item \texttt{measurement.main\_reference}: link to the original publication if .source is a compilation publication.
    \item \texttt{measurement.note}: added notes informing about a publication's assumptions or derivation of the value.
    \item \texttt{measurement.name}: the list of aliases of the object.
\end{itemize}

Due to the differences in availability of metadata, other properties exist depending on the type of measurement. 
Certain measurement types also require a specifier upon retrieval (e.g. the emission line for a line width).

\paragraph{Lag.} The Lag class represents a single published reverberation 
mapping measurement together with all available metadata describing the 
observing campaign. It is called by \texttt{.lag(line)} and contains asymmetric 
errors accessed by \texttt{lag.error\_plus} and \texttt{lag.error\_minus} and
depending on availability \texttt{.baseline, .cadence, .epochs, .snr, 
.grade, .method}. Grade refers to quality flags assigned by the publication
and method to the employed recovery algorithm. With \texttt{lag.luminosity($\lambda$)}
the corresponding continuum luminosity measurement from the same publication is 
retrieved automatically whenever a unique match exists.

\paragraph{Line-width.} The line-width class is called by \texttt{.linewidth(line, type, spec\_type)}
and contains the error of the measurement accessed by \texttt{linewidth.error}.
For type it is possible to choose between \texttt{FWHM} and \texttt{line dispersion}
measurements as well as \texttt{rms} and \texttt{mean} for the type of spectrum the
measurement is taken from.

\paragraph{Luminosity.} The luminosity class is called using 
\texttt{.luminosity($\lambda$)} with $\lambda$ in~\AA~and it contains the 
error of the measurement which is accessed by \texttt{luminosity.error}
and the cosmology model used to calculate the measurement (\texttt{luminosity.cosmology}).
With \texttt{luminosity.convert(new\_cosmology)} it is possible to calculate the
luminosity if a different cosmology model is used compared to the one from the publication.
The argument of the function has to be an \texttt{astropy.cosmology} object. 
With \texttt{luminosity.lag(line)}
the luminosity measurement can be directly matched to the lag measurement
from the same publication, if available. This removes the need for manual 
matching of luminosities and lags across publications.

\paragraph{Mass.} Called by \texttt{.mass(line)} this class contains the assumed virial
factor and error for the mass calculation accessed by \texttt{mass.virial\_factor}
and \texttt{mass.virial\_factor\_error} as-well as the asymmetric uncertainties
\texttt{mass.error\_plus} and \texttt{mass.error\_minus}. 
Because the mass depends on the adopted spectrum and line-width type, these can be inspected via \texttt{mass.spectrum\_type} and \texttt{mass.linewidth\_type}.

\paragraph{Virial product.} Called by \texttt{.vp(line)} it is analogous to the mass since it is derived by dividing the mass by the virial factor.

\paragraph{Redshift.} Is called with \texttt{.redshift} and only
contains the default properties.

\paragraph{RA, Dec, Position.} Are called by \texttt{.ra}, \texttt{.dec}
and \texttt{.position} respectively and contain only the default
properties.

\paragraph{Distance.} Is called by \texttt{.distance(method, *cosmology)} where the
method refers to the method used in the distance ladder, such as
the luminosity distance and group-averaged distance. It only contains the
error \texttt{distance.error} as a property. If it is called without a method
and a cosmology model added instead, it returns the distance based on the redshift
directly as a measurement.

\subsubsection{MeasurementCollection}
\label{app:measurementcollection}

A \texttt{MeasurementCollection} stores multiple measurements of the same observable. 
Most AGNs have several published lag or luminosity measurements, making it 
necessary to distinguish between the collection itself and the individual Measurement 
objects it contains. Measurement collections behave similarly to Python lists and 
therefore support iteration and length operations while also providing additional 
methods for filtering, matching and combining measurements.

The individual measurements have to be accessed by iterating through the 
collection. For example 
\begin{verbatim}
agn = db.get('NGC 5548')
lag_collection = agn.lag('H_beta')
for lag in lag_collection:
    print(lag.value)
\end{verbatim}  
returns the individual lag values of each publication in the database for
NGC 5548.

Besides iteration, every measurement collection provides several utility
functions for selecting or combining measurements.

\paragraph{filter(**kwargs)} allows the user to select measurements based on attributes of the \texttt{Measurement} class. 
To filter for lag values of a specific publication which are marked as unproblematic,
\begin{verbatim}
lags = agn.lag("H_beta").filter(
                source="Bentz2013",
                problematic=False)
\end{verbatim}
can be used. 

\paragraph{match(measurement)} returns the corresponding measurement 
describing the same observing campaign whenever a unique match exists. 
This can be used to select the corresponding luminosity measurement to 
the lag measurement of the same publication.
\begin{verbatim}
for lag in lags:
    luminosity = agn.luminosity(5100).match(lag)
\end{verbatim}

In case of mass and virial product collections, it is possible to
combine the various measurements to derive a single mass for the AGN.
By calling \texttt{combine(spectra\_type, linewidth\_type)} it can
be chosen what type of measurements to include. The combined value is 
computed from the envelope of all selected measurements, where the central 
value is taken as the midpoint of the full allowed range and the uncertainty
corresponds to half of the total envelope width.
\begin{verbatim}
mass = agn.mass('H_beta').combine(
                        spectra_type = 'rms', 
                        linewidth_type = 'FWHM')
print(mass.value, mass.error)
\end{verbatim}

\subsection{PublicationView}

Many analyses require access to all measurements reported by a 
specific publication, rather than all measurements belonging to
a single AGN. For this purpose the \texttt{PublicationView} class
provides a publication-centred view of the database, while retaining
the object-oriented structure of the individual measurements.
It is obtained by

\begin{verbatim}
view = db.publication_view(source)
\end{verbatim}

where \texttt{source} is a reference corresponding to one of the
publications included in the database. Since the database
back-end functions with links to the publications, they have
to be provided as such using \texttt{qrm.link\_finder(short\_reference)}
where short\_reference refers to author year style, i.e. Shen2024.

The returned object contains all measurements published by the selected
reference, grouped by AGN and measurement type. This allows users
to conveniently reproduce literature samples or compare the measurements
of a single publication to values from other studies.

The AGN contained in the publication can be obtained by
\begin{verbatim}
for agn in view.agns():
    print(agn.name)
\end{verbatim}
while all measurements belonging to a particular AGN can be accessed
using
\begin{verbatim}
measurements = view.measurements_by_agn(agn)
\end{verbatim}
where the argument agn refers to a common identifier. The returned
dictionary contains all available lags, luminosities, line-widths,
black hole masses, virial products and distance measurements reported
by the publication for the selected AGN as part of the 
\texttt{MeasurementCollection} class.

\subsection{Scout}
\label{app:scout}
The \texttt{scout()} routine provides the framework to estimate
ICCF performance developed in this work (see Appendix~\ref{app:RM-Scout}
for further information). The simulation is started using
\begin{verbatim}
result = qrm.scout(luminosity,
                z, baseline,
                cadence, sn)
\end{verbatim}
where \texttt{luminosity} is the continuum luminosity in erg/s,
\texttt{baseline} the total observing baseline,
\texttt{cadence} the sampling cadence and \texttt{sn} denotes the typical
S/N of the planned observations.

By default, the expected lag is calculated from this work's calibrated
$R$-$L$ relation of H$\beta$. Alternative linear relations can be supplied
through the \texttt{relation} argument together with the corresponding
model parameters $\alpha$ and $\beta$.

The returned \texttt{SimulationResult} object stores both the input parameters and all recovered lag measurements obtained during the
simulation. Summary statistics such as the recovered lag, 16th/84th
percentile of the distribution, outlier fraction and success rate are
calculated automatically and are available as class properties.

Several visualisation routines are implemented directly as methods
of the result object. For example,
\begin{verbatim}
result.view()
\end{verbatim}

produces a summary figure containing a simulated and sampled 
light curve, recovered lags on the $R$-$L$ plane and bias
distribution. For more information about the interpretation
of these results, refer to Appendix~\ref{app:RM-Scout}. 
Individual components of the simulation can also be visualised
separately using dedicated plotting functions such as

\begin{verbatim}
result.iccf(index)
result.bias_histogram()
result.rl_plane()
\end{verbatim}
If \texttt{index} is provided, the ICCF of the corresponding realisation is shown; otherwise, a stacked view of all recovered ICCFs is displayed. 
This allows users to inspect individual aspects of the simulation in greater detail.

\section{Description of the full simulation results table}
\label{app:simtable}
 
Table~\ref{tab:sim_summary} lists the simulation results for the full
H$\beta$ sample under the assumption of $\alpha_{\rm ref} = 0.5$ and
flag thresholds 0.5/1.5. From left to right, the columns display:
object name;
campaign source code (1 = \citealt{McDougall2025}, 2 = \citealt{Hu2025}, 3 = \citealt{Woo2024},
4 = \citealt{Bentz2013}, 5 = \citealt{Shen2024}, 6 = \citealt{Hu2021}, 7 = \citealt{Grier2017});
reported lag $\tau_{\rm rep}$ (days);
expected lag $\tau_{\rm exp}$ (days);
retrieved lag $\tau_{\rm ret}$ with asymmetric uncertainties (days)
defined by the 16th and 84th percentile of the distribution;
simulated coverage fraction $f_{\rm sim.~cover}$ defined as the ratio
of simulated observations and planned observations before accounting
for seasonal gaps;
redshift $z$;
campaign baseline $T$ (days);
nominal cadence $\delta T$ (days);
effective cadence $\delta T_{\rm sim}$ used in the simulation (days);
number of reported epochs $N_{\rm rep}$;
number of simulated epochs $N_{\rm sim}$;
estimated S/N;
simulated S/N;
continuum luminosity $L_{5100}$ ($\mathrm{erg\,s^{-1}}$);
quality flag (0-3).
A value of -1 (e.g. the $\delta T$ entry of 3C 390.3) indicates that the value was not available or, in the case of the S/N, could not be estimated.
 
The full table (248 rows) is available in machine-readable form at
the CDS and as part of the database distribution. Analogous tables
for the discussed simulation setups (different reference slopes and
flagging thresholds) are available likewise.
A representative excerpt for well-known objects is shown below.
 
\begin{table*}[htbp]
  \caption{Excerpt from the simulation results table for the H$\beta$
           $R$-$L$ sample. }
  \label{tab:sim_summary}
  \centering
  \small
  \begin{tabular}{lcccccccccccccccc}
    \hline\hline
    Name & C &  $\tau_{\rm rep}$ & $\tau_{\rm exp}$ & $\tau_{\rm ret}$ & $f_{\rm sim.~cover}$ & $z$ & $T$ & $\delta T$ & $\delta T_{\rm sim}$ & $N_{\rm rep}$ & $N_{\rm sim}$ & S/N & S/N$_{\rm sim}$ & $L_{5100}$ & Flag \\
         &   & (d)  & (d) & (d) & & & (d) & (d) & (d) & & & & & ($10^{44}$erg~s$^{-1}$)\\
    \hline
    Mrk 1501 & 3 & 11.7& 34.9& $34.6    ^{+15.2}_{-14.9}$ &     0.52 & 0.09     & 1741 &  15 & 15 & 38 & 61 &41 & 41 & 1.22 &     2 \\
    Mrk 1014 & 2 & 95.9 & 56.4 & $56.9^{+21.3}_{-16.9}$ & 0.55 & 0.16 & 2018 & 6.0 & 6.0 & 164 & 184 & 145 & 145 & 3.18 & 2 \\
    3C 120 & 4 & 25.9 & 27.2 &  $26.6^{+1.9}_{-4.2}$ & 1.0 & 0.03 &     128 & 0.99 & 0.99 & 84 & 130 & 40 & 40 & 0.74 &     0 \\
    3C 390.3 & 4 & 46.4 &       51.9 &  $11.6^{+17.3}_{-21.6}$ & 1.0 & 0.06 &       83.0 & -1.0 & 3.32 & 25 & 25 & -1 & 100 & 2.69 &        3 \\
    \hline
  \end{tabular}
  \tablefoot{See Appendix~\ref{app:simtable} for a
           description of all columns. All lags are in rest-frame days.}
\end{table*}

\section{Fits for each subsample}
\label{app:fits}

\begin{figure*}
    \centering
    \includegraphics[width=0.195\linewidth]{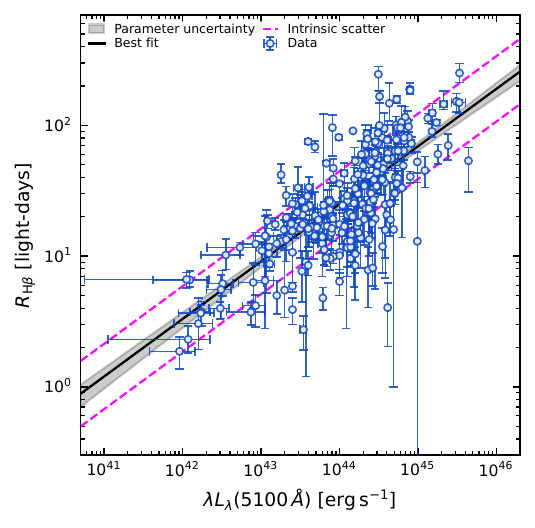}
    \includegraphics[width=0.195\linewidth]{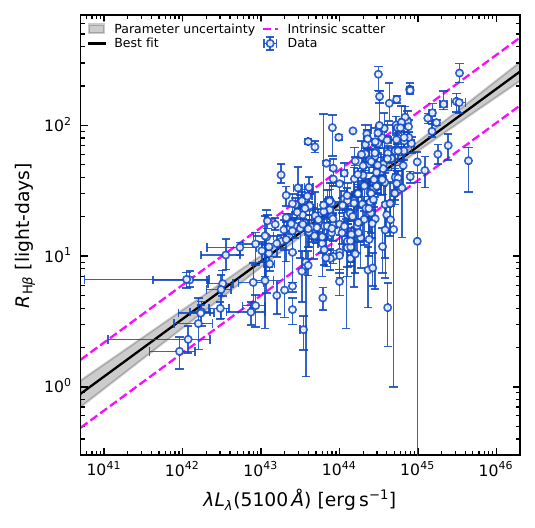}
    \includegraphics[width=0.195\linewidth]{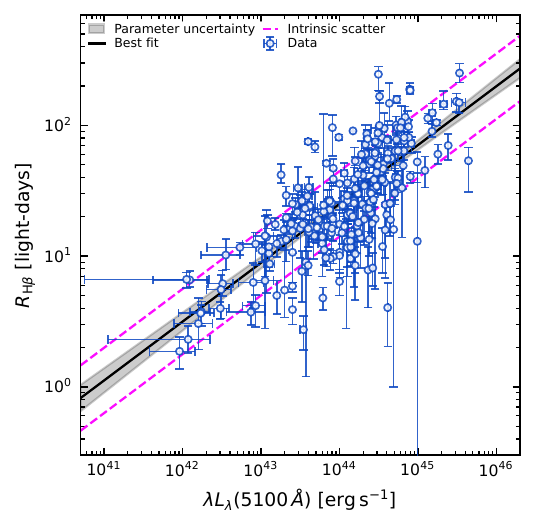}
    \includegraphics[width=0.195\linewidth]{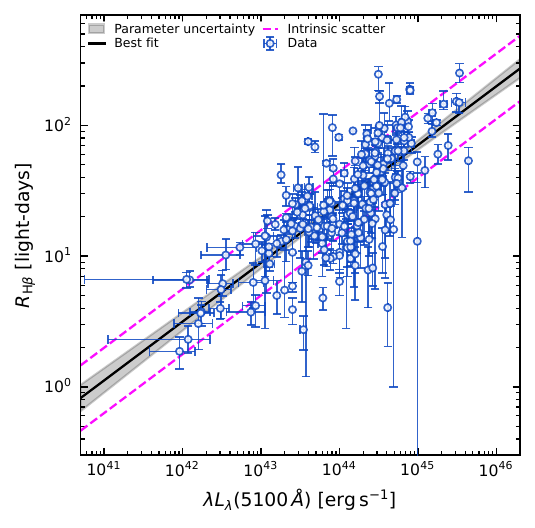}
    \includegraphics[width=0.195\linewidth]{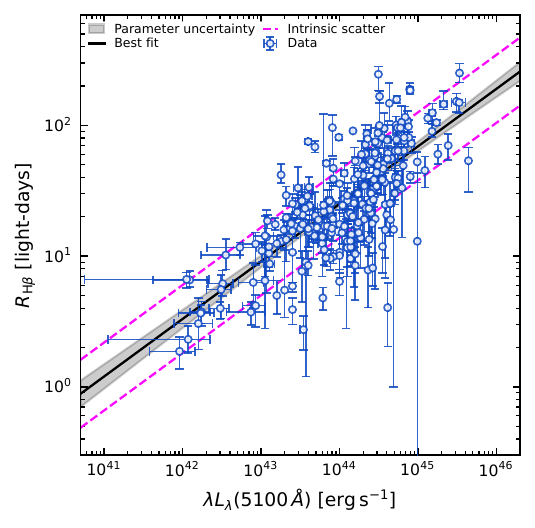}
    \includegraphics[width=0.195\linewidth]{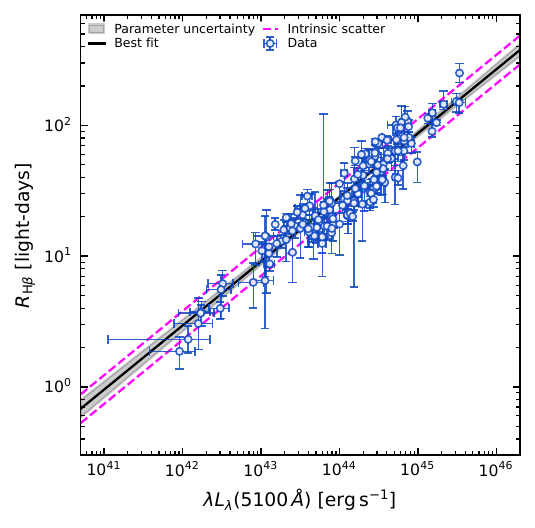}
    \includegraphics[width=0.195\linewidth]{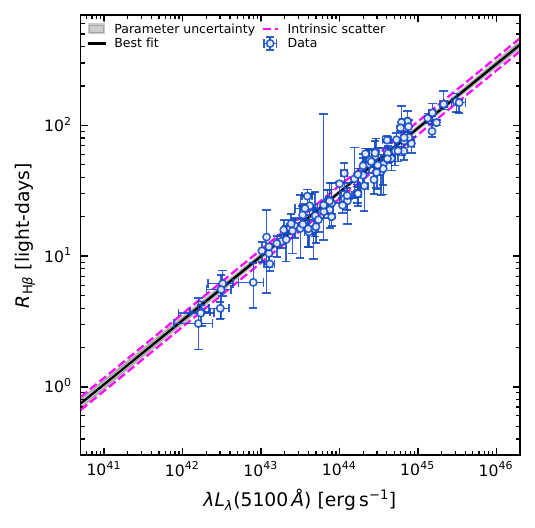}
    \includegraphics[width=0.195\linewidth]{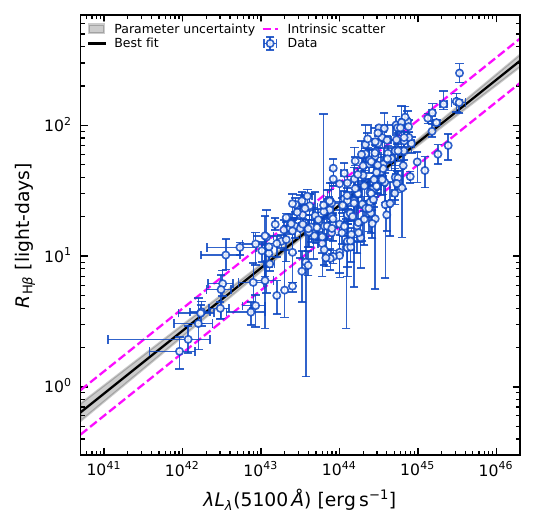}
    \includegraphics[width=0.195\linewidth]{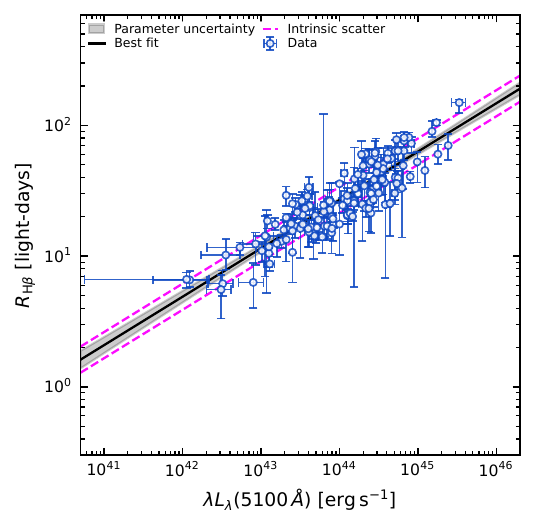}
    \includegraphics[width=0.195\linewidth]{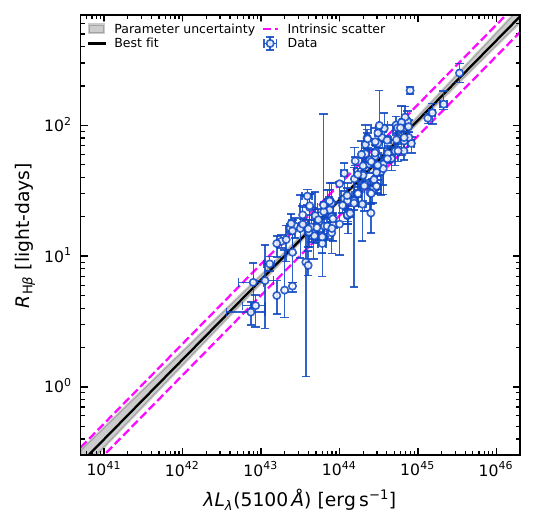}
    \includegraphics[width=0.195\linewidth]{figures/fits/fit_cat_0.pdf}
    \includegraphics[width=0.195\linewidth]{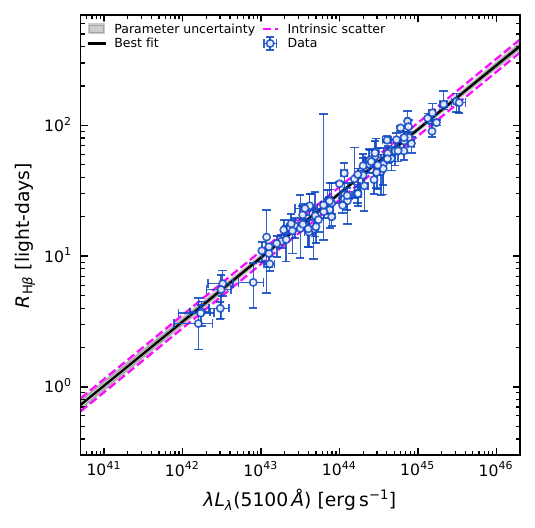}
    \includegraphics[width=0.195\linewidth]{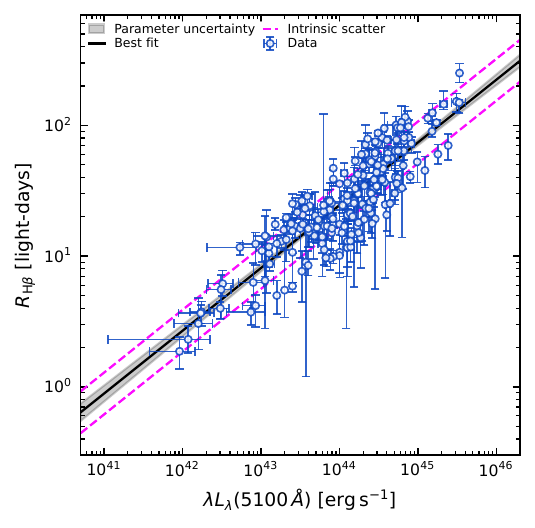}
    \includegraphics[width=0.195\linewidth]{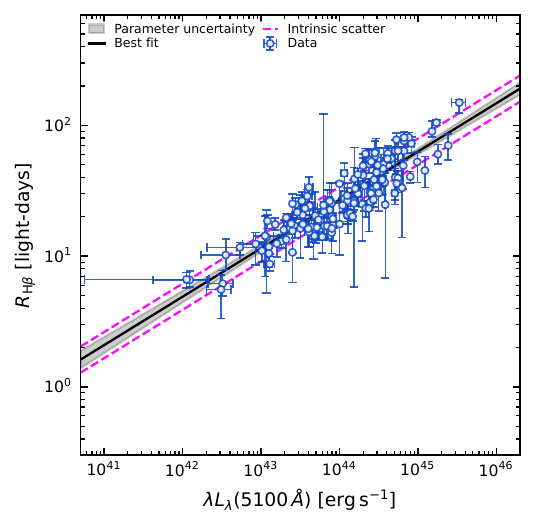}
    \includegraphics[width=0.195\linewidth]{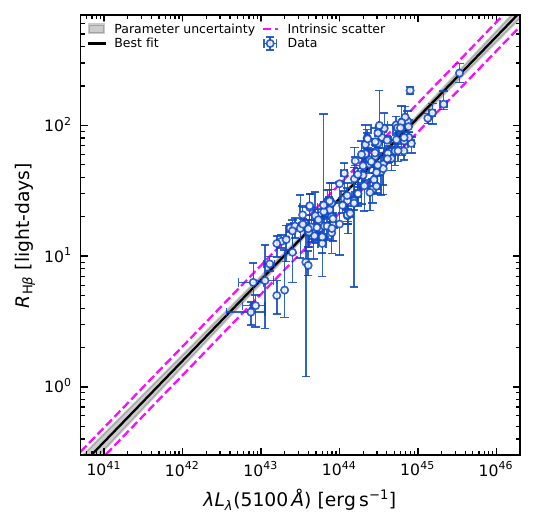}
    \caption{Best-fits of the H$\beta$ $R$-$L$ relation for the different categories and reference models. First row shows the fits when including Cat.\ 0+1+2, second row Cat.\ 0+1 and third row Cat.\ 0. In each row the panels show from left to right: (1) Fit for the threshold 0.5/1.5 when using $\alpha_{\rm ref} = 0.5$, (2) Fit for the threshold 0.7/1.3 when using $\alpha_{\rm ref} = 0.5$, (3) Fit for the threshold 0.3/1.7 when using $\alpha_{\rm ref} = 0.5$,
    (4) Fit for the threshold 0.5/1.5 when using $\alpha_{\rm ref} = \frac{1}{3}$, (5) Fit for the threshold 0.5/1.5 when using $\alpha_{\rm ref} = 0.7$.}
    \label{fig:fit_matrix}
\end{figure*}

\begin{figure*}
    \centering
    \includegraphics[width=0.195\linewidth]{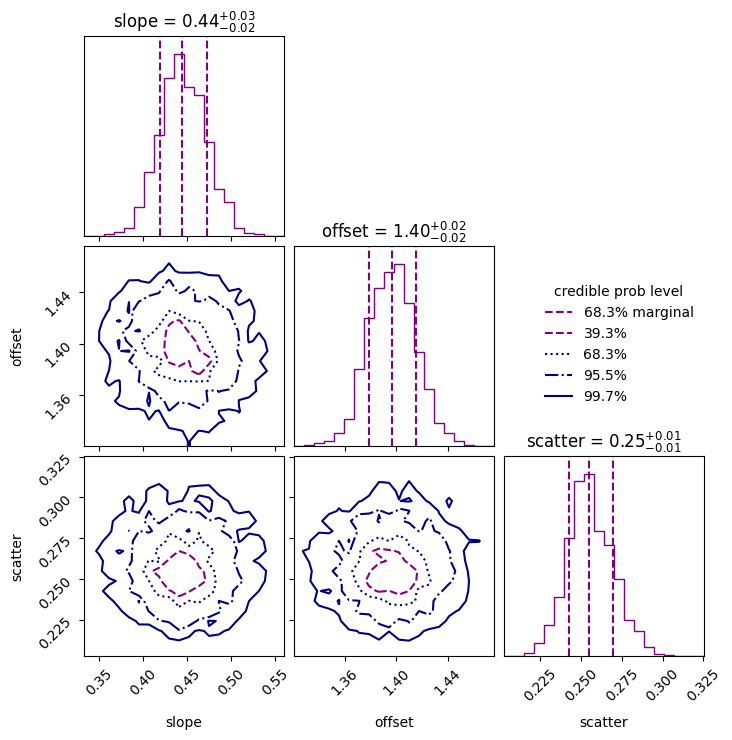}
    \includegraphics[width=0.195\linewidth]{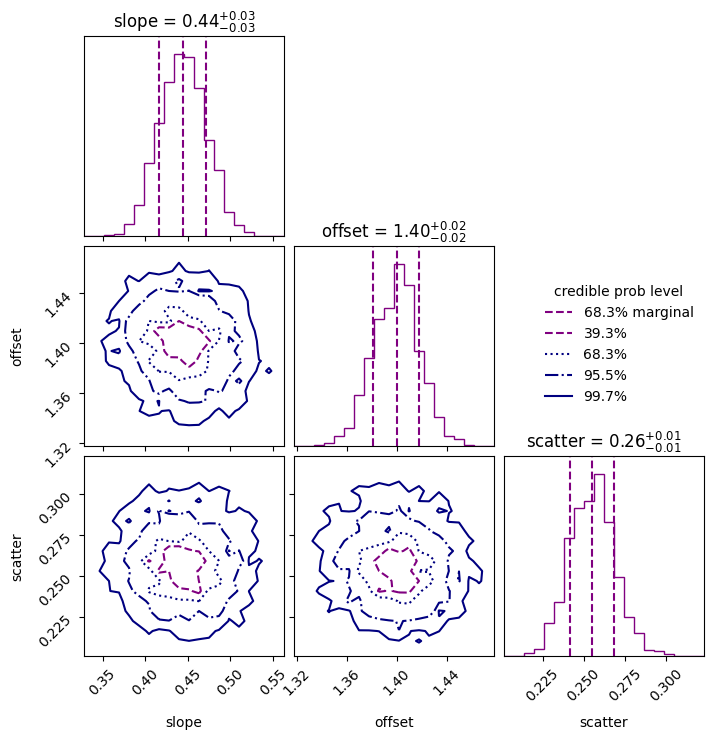}
    \includegraphics[width=0.195\linewidth]{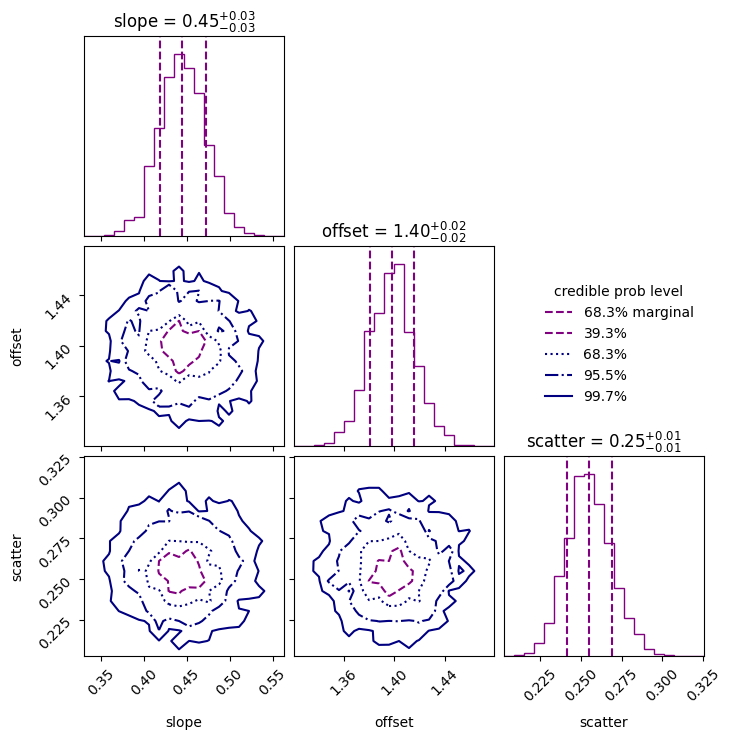}
    \includegraphics[width=0.195\linewidth]{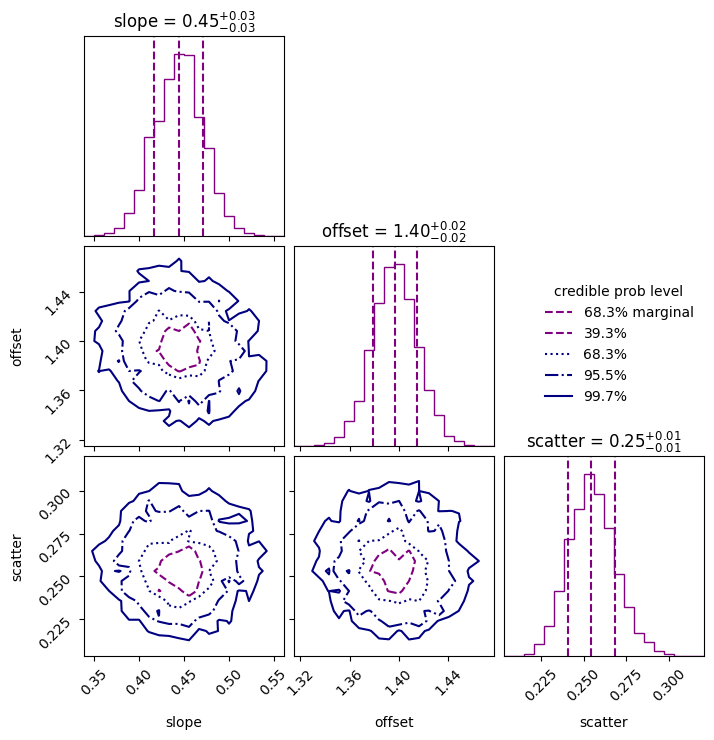}
    \includegraphics[width=0.195\linewidth]{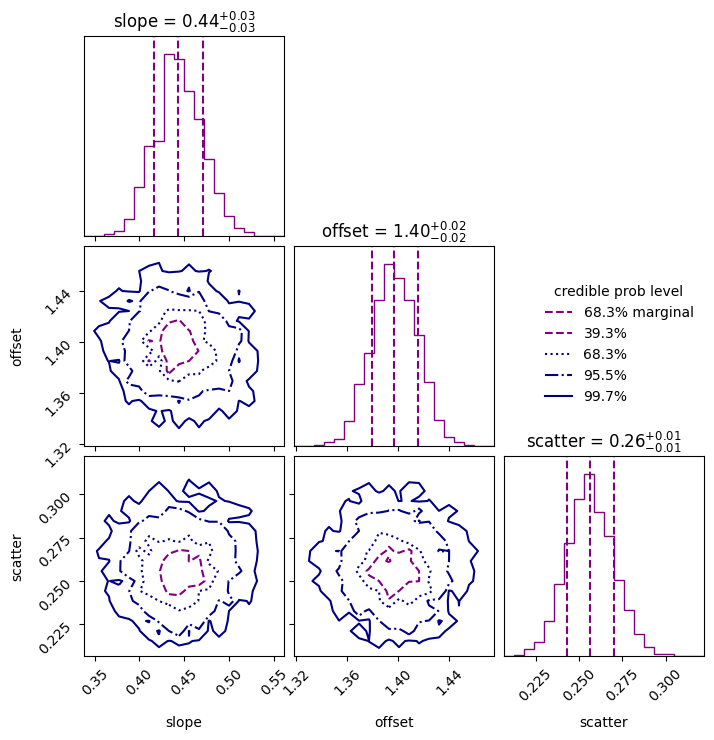}
    \includegraphics[width=0.195\linewidth]{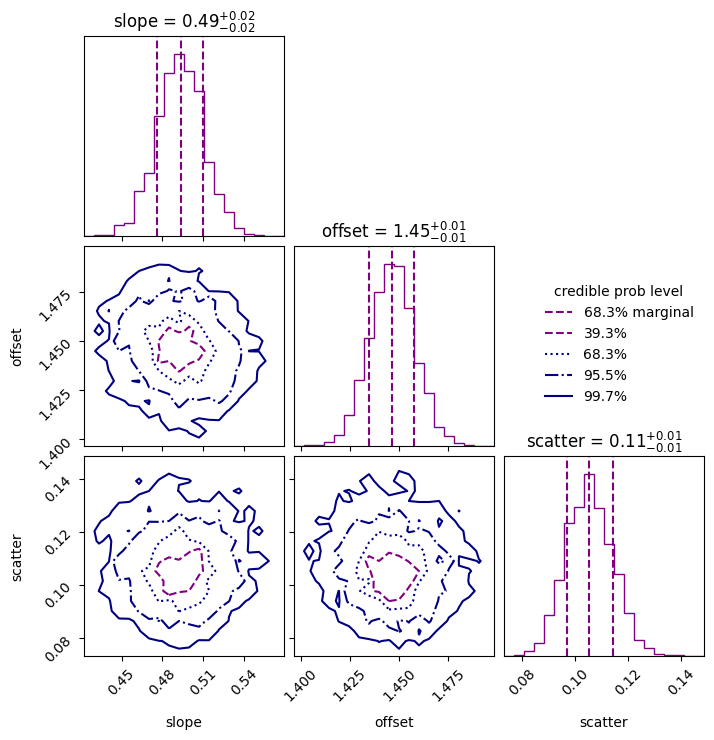}
    \includegraphics[width=0.195\linewidth]{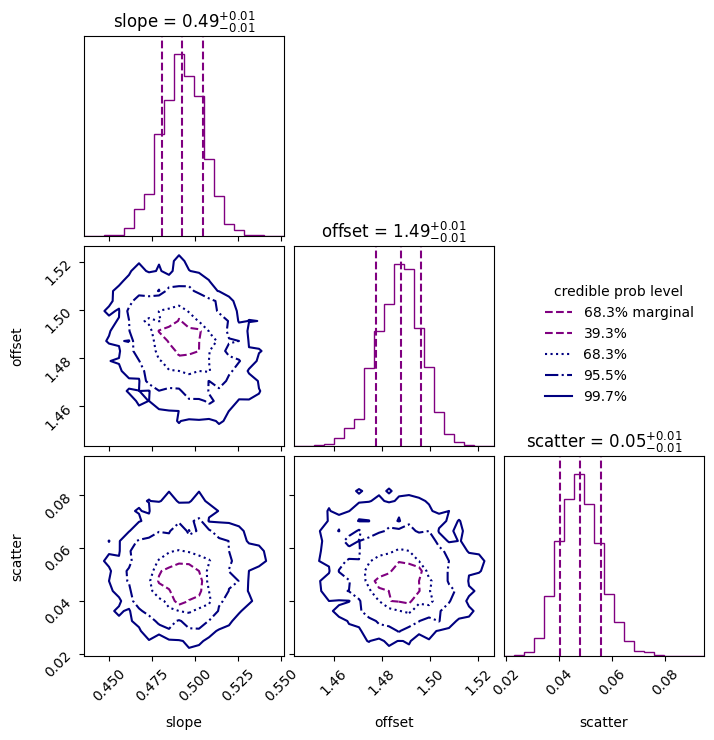}
    \includegraphics[width=0.195\linewidth]{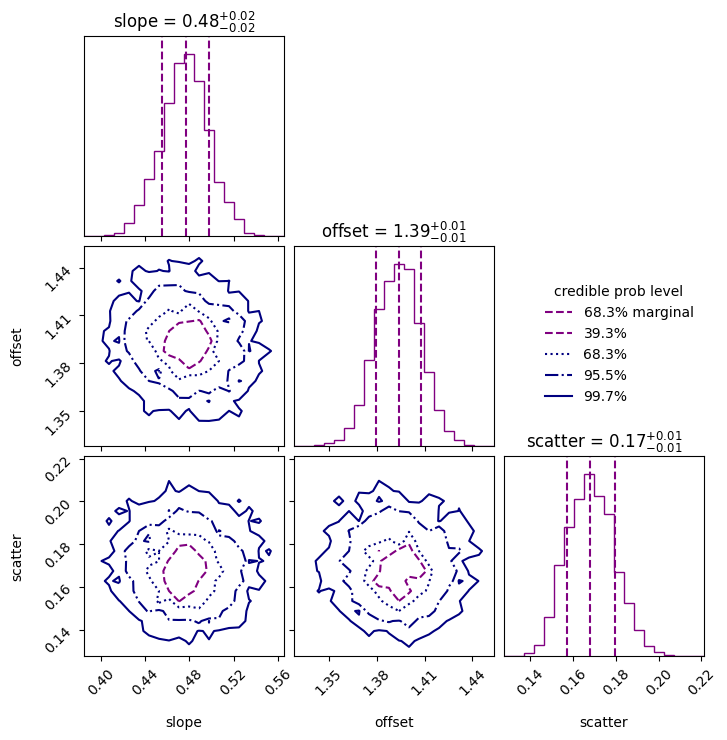}
    \includegraphics[width=0.195\linewidth]{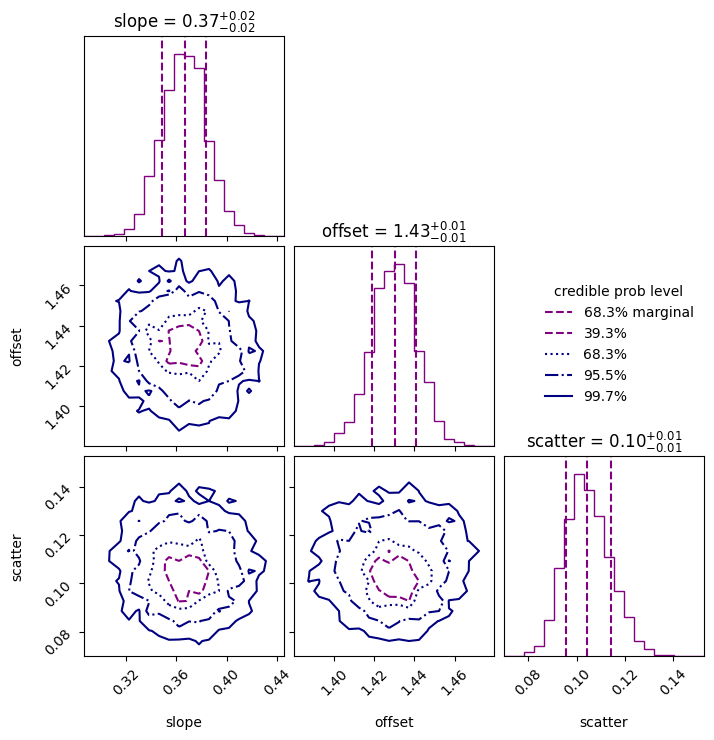}
    \includegraphics[width=0.195\linewidth]{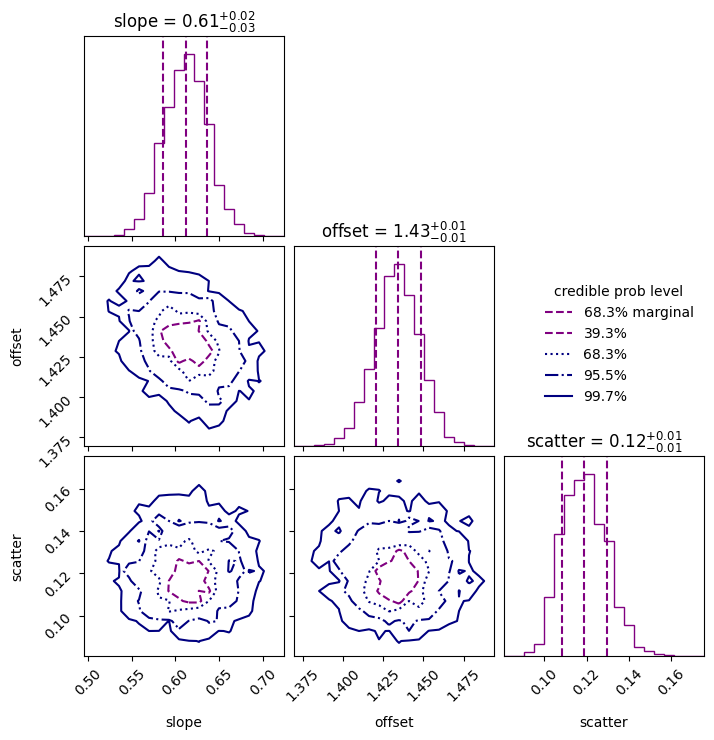}
    \includegraphics[width=0.195\linewidth]{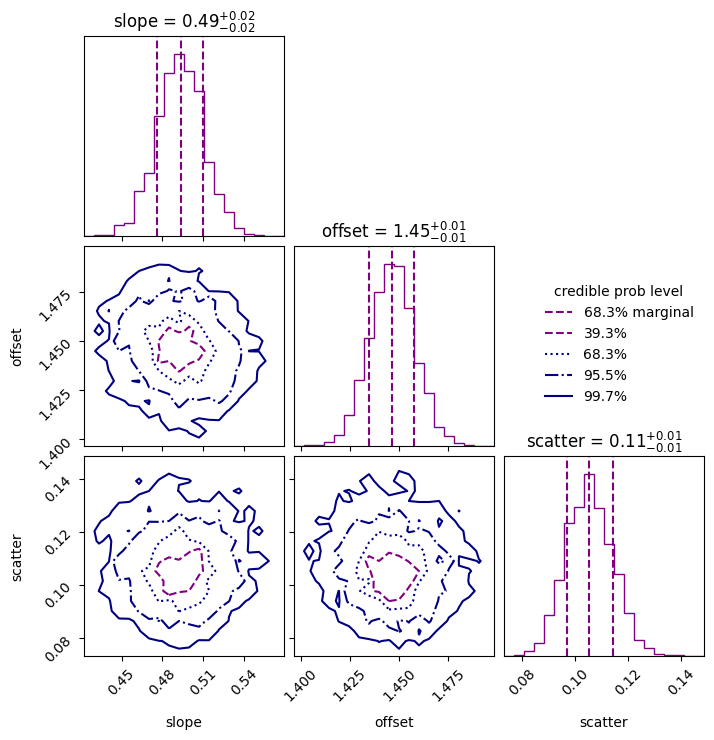}
    \includegraphics[width=0.195\linewidth]{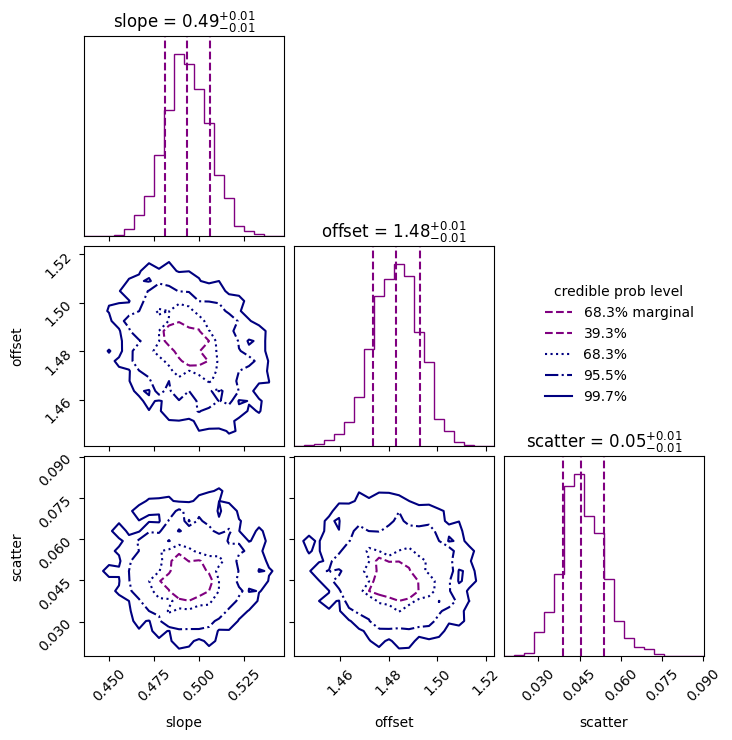}
    \includegraphics[width=0.195\linewidth]{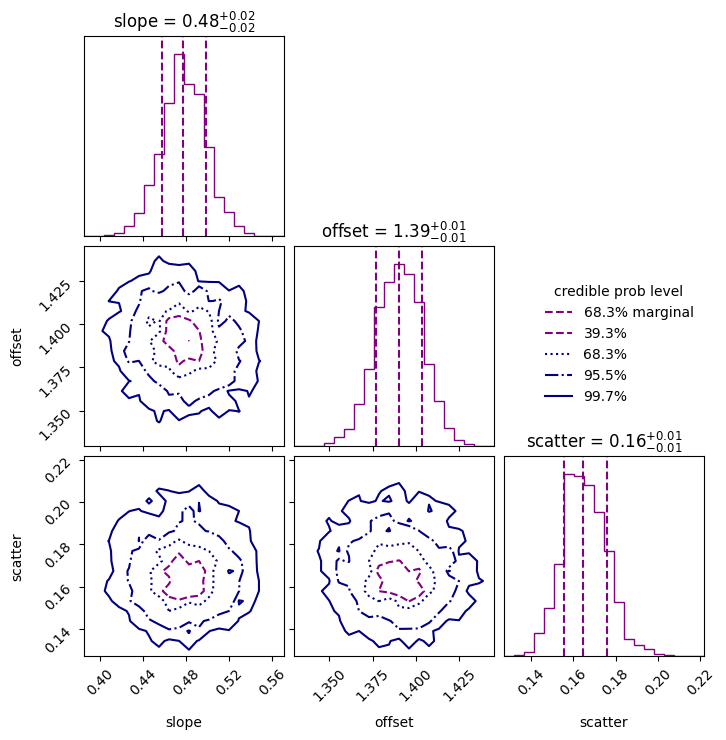}
    \includegraphics[width=0.195\linewidth]{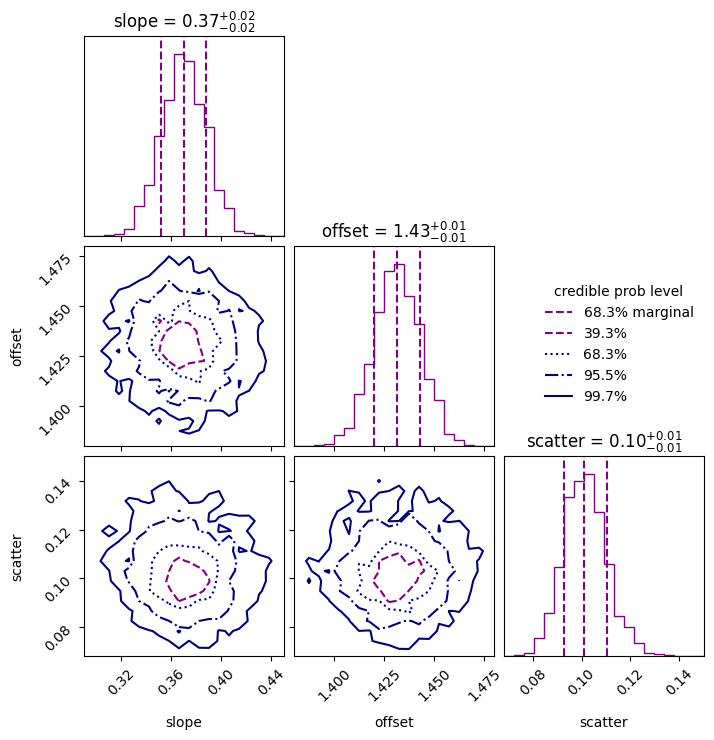}
    \includegraphics[width=0.195\linewidth]{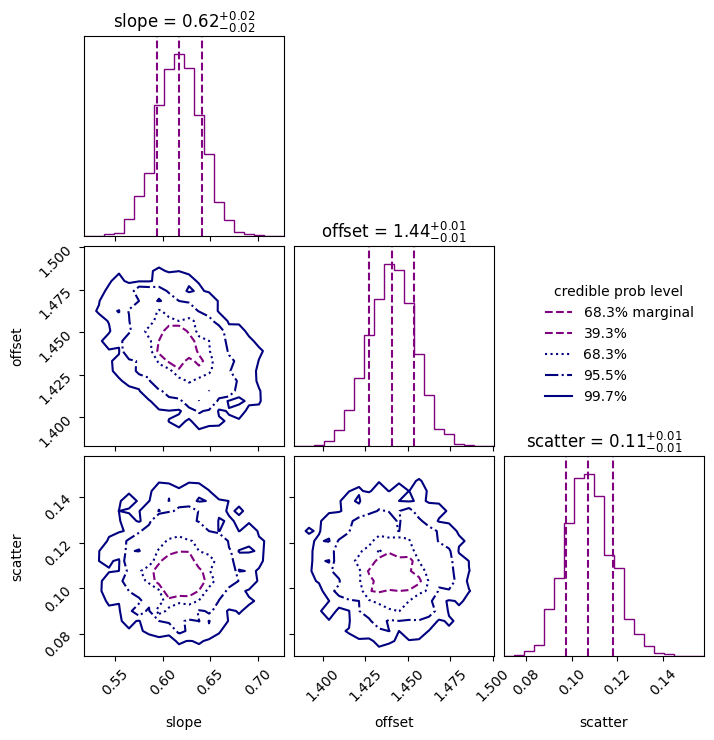}
    \caption{Posterior corner-plots of the Best-fit H$\beta$ $R$-$L$ relation. First row shows the posteriors when fitting Cat.\ 0+1+2, second row Cat.\ 0+1, and third row Cat.\ 0. In each row the panels show from left to right: (1) Posterior for the threshold 0.5/1.5 when using $\alpha_{\rm ref} = 0.5$, (2) Posterior for the threshold 0.7/1.3 when using $\alpha_{\rm ref} = 0.5$, (3) Posterior for the threshold 0.3/1.7 when using $\alpha_{\rm ref} = 0.5$,
    (4) Posterior for the threshold 0.5/1.5 when using $\alpha_{\rm ref} = \frac{1}{3}$, (5) Posterior for the threshold 0.5/1.5 when using $\alpha_{\rm ref} = 0.7$.}
    \label{fig:posterior_matrix}
\end{figure*}

Figure~\ref{fig:fit_matrix} shows the individual fits
of the H$\beta$ $R$-$L$ relation for each subsample 
after the exclusion of flagged objects.
The rows show from top to bottom: Best fit including
Cat.\ 0+1+2, Cat.\ 0+1, Cat.\ 0.
The columns display the best fit for the threshold: 
(1) 0.5/1.5 when using $\alpha_{\rm ref} = 0.5$, 
(2) 0.7/1.3 when using $\alpha_{\rm ref} = 0.5$, 
(3) 0.3/1.7 when using $\alpha_{\rm ref} = 0.5$,
(4) 0.5/1.5 when using $\alpha_{\rm ref} = \frac{1}{3}$, 
(5) 0.5/1.5 when using $\alpha_{\rm ref} = 0.7$.
Figure~\ref{fig:posterior_matrix} shows the posterior
distributions of the fits in the same row/column style
as described above.

\begin{figure*}
    \centering
    \includegraphics[width=0.245\linewidth]{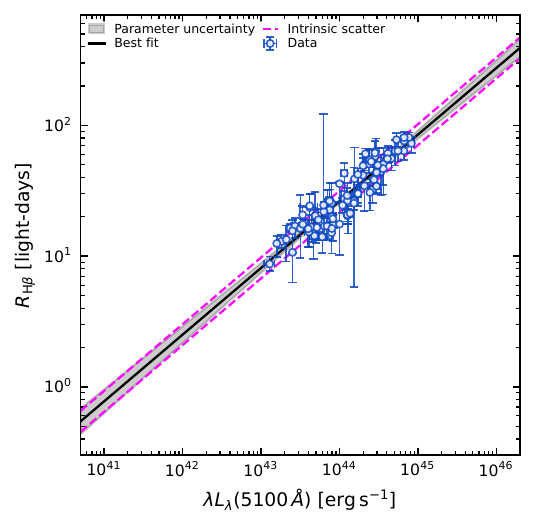}
    \includegraphics[width=0.245\linewidth]{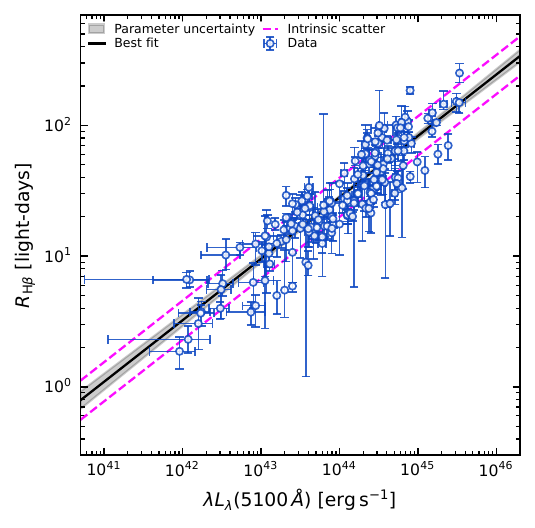}
    \includegraphics[width=0.245\linewidth]{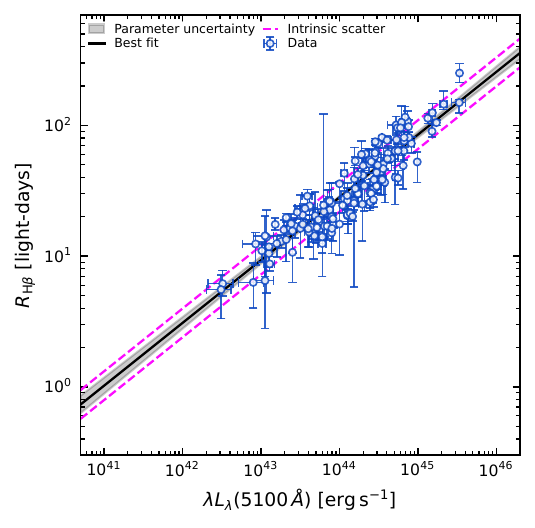}
    \includegraphics[width=0.245\linewidth]{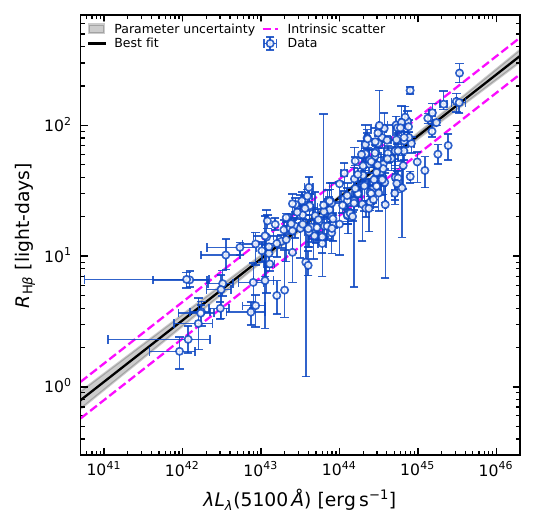}
    \includegraphics[width=0.245\linewidth]{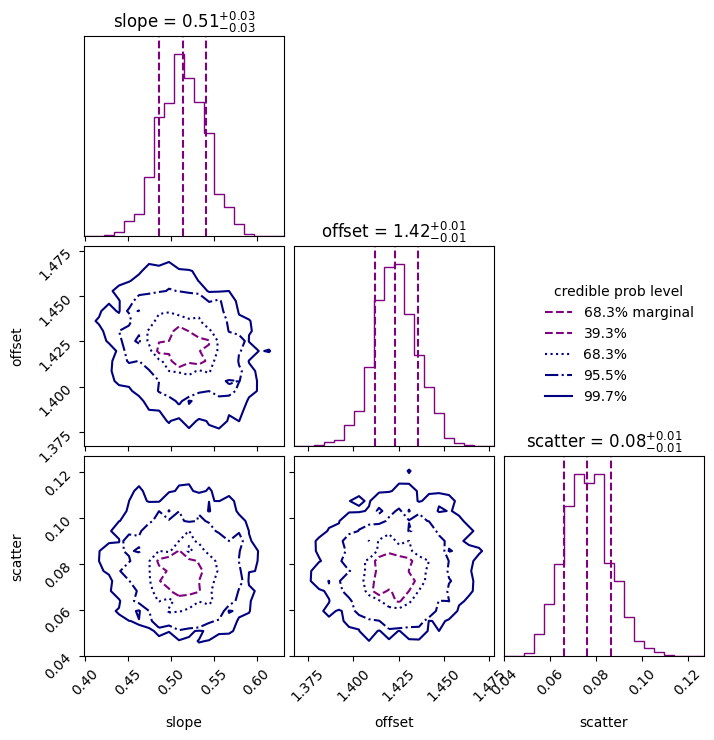}
    \includegraphics[width=0.245\linewidth]{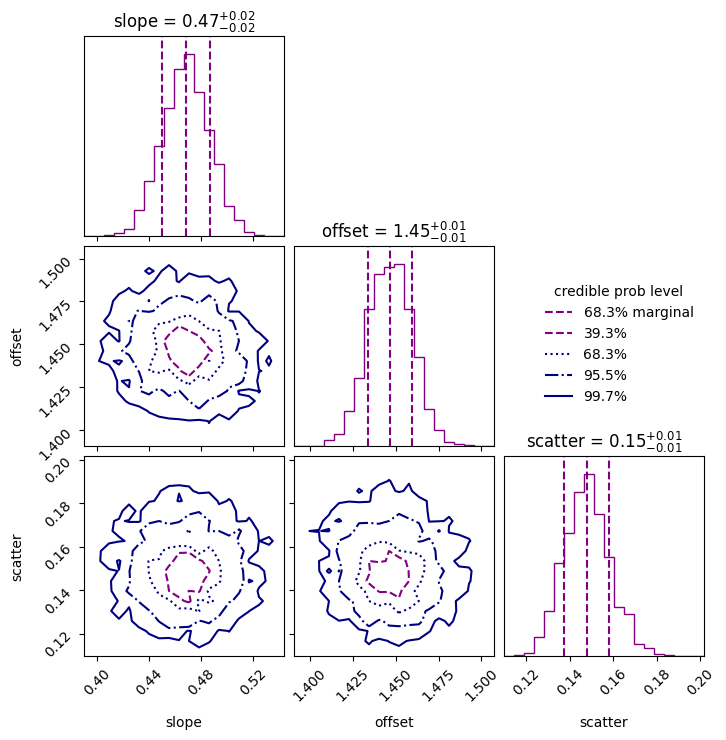}
    \includegraphics[width=0.245\linewidth]{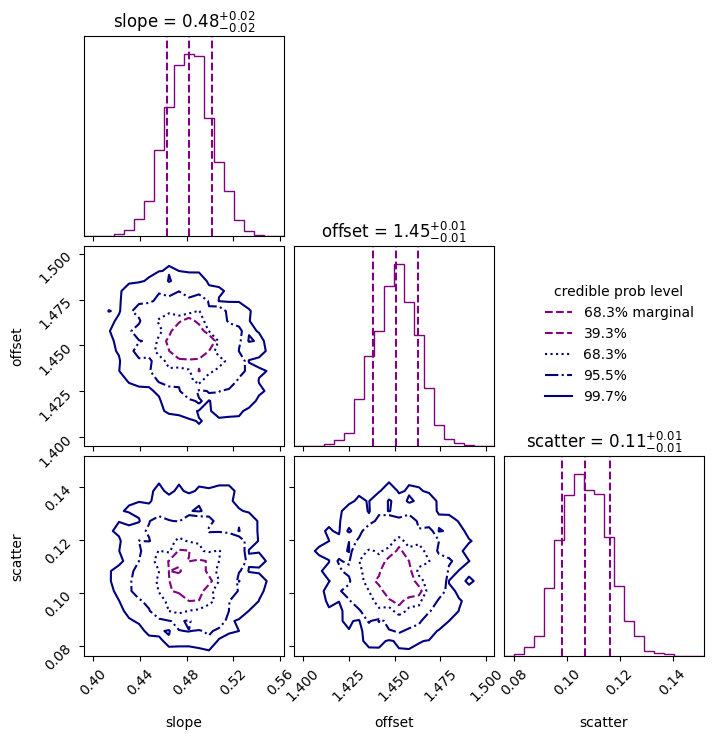}
    \includegraphics[width=0.245\linewidth]{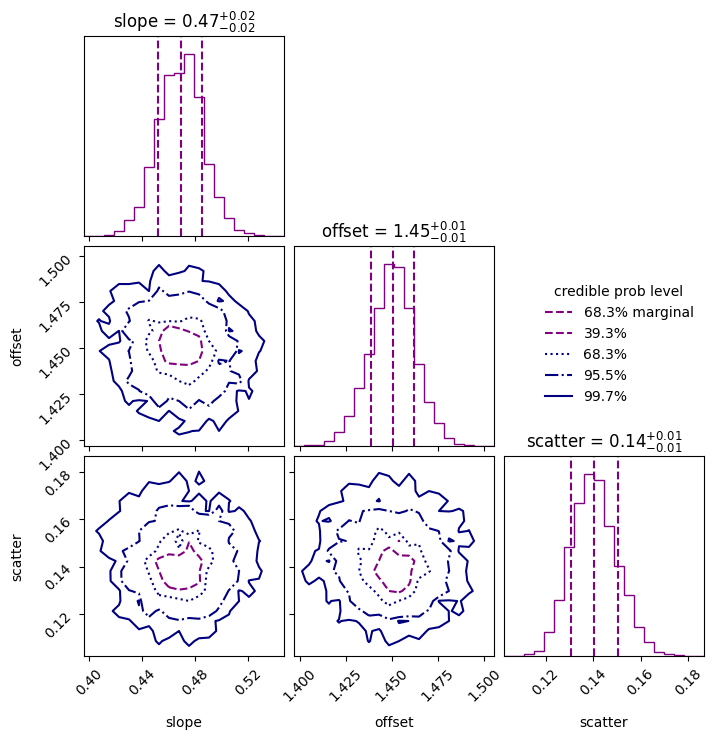}
    \caption{\textit{Top row}: Best fits for the model corrected samples.
            Columns are from left to right: (1) Sample 1; (2) Sample 2;
            (3) Sample 3; (4) Sample 4.
            \textit{Bottom row}: Corresponding posterior corner-plots to the
            fits in the row above.}
    \label{fig:intersect_matrix}
\end{figure*}

Figure~\ref{fig:intersect_matrix} shows the fits (top row) and
posterior distributions (bottom row) of the defined subsamples
in Sect.~\ref{sec:model-bias-correction} correcting
for the model bias. Columns are from left to right:
(1) Sample 1; (2) Sample 2; (3) Sample 3; (4) Sample 4.

The figures are available in full size on GitHub.

\section{The \texttt{scout} simulation tool}
\label{app:RM-Scout}
Scout as part of the Python package \texttt{pyRMTools} is developed to estimate the
adequacy of a set of RM observational parameters given a 
continuum luminosity, redshift, baseline, cadence
and S/N. See Appendix~\ref{app:scout} for the documentation.

After providing these input parameters to the program,
the lag $\tau_{\rm est}$ of the object is calculated in
observer frame using this work's calibration as a 
reference relation or user chosen parameters.

\begin{equation}
    \label{eq:est_lag_rm_scout}
  \log\!\left(\tau_{\rm est}\right) = 1.45 + 0.48\,
  \log\!\left(\frac{L_{5100}}{10^{44}\,\mathrm{erg\,s^{-1}}}\right)
  + \log(1+z).
\end{equation}

The variability amplitude (rms) of the light curve was 
estimated using Eq.~\ref{eq:sf} and rms = SF/$\sqrt{2}$.
To mimic the user-specified observing parameters, the light
curve was sampled on an evenly spaced grid and seasonal gaps were inserted as described in Sect.~\ref{sec:obs_sampling}. Next, Gaussian white noise was added to the sampled
light curves with a standard deviation set by the S/N. One thousand such light-curve pairs were generated and the ICCF was applied to each to recover the lag. The ICCF parameters were set to search for a lag between $-0.3\cdot T$ and $0.5\cdot T$ with a spacing of $0.8\cdot\delta T$.

\begin{figure*}
    \centering
    \includegraphics[width=1\linewidth]{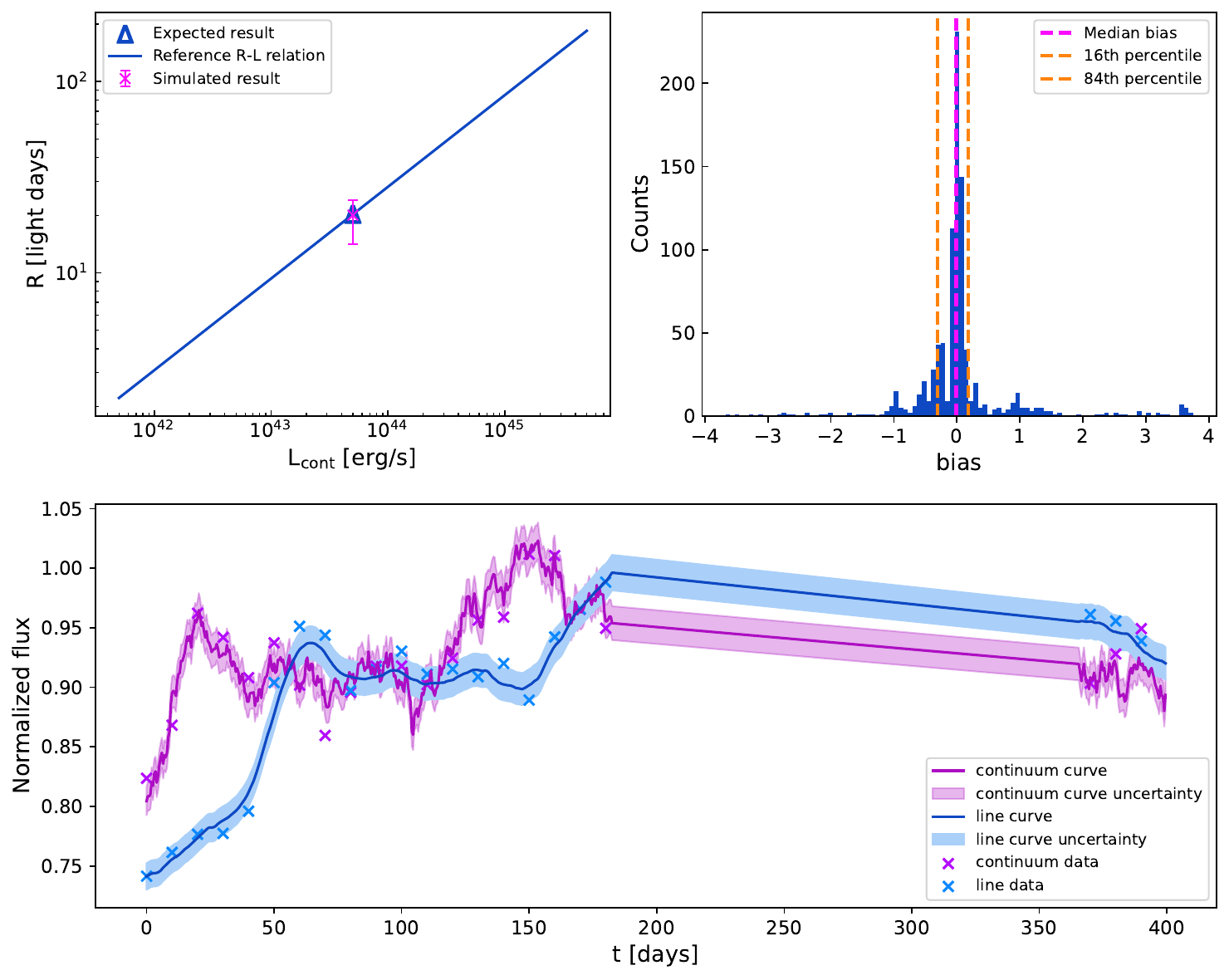}
    \caption{Output figure of the RM-Scout program for
    $L_{\rm cont}=5\times10^{43}$erg/s, $z=1$, $T = 400$ days, 
    $\delta T = 10$ days, S/N $= 67$.
    \textit{Top left}: Expected rest-frame lag and recovered
    rest-frame lag compared with the underlying assumed
    reference $R$-$L$ relation.
    \textit{Top right}: Bias distribution of the recovered
    lags with median and 16th/84th percentile marked as
    dashed lines.
    \textit{Bottom}: Continuum and line light curve with
    the sampled data according to the observing conditions
    and uncertainties in the light curves marked as shaded
    areas based on the S/N.}
    \label{fig:figure_rm_scout}
\end{figure*}

The function outputs a figure (see e.g. Fig.~\ref{fig:figure_rm_scout})
which visualises important metrics of the ICCF analysis.
The top left panel shows the expected rest-frame lag and 
recovered rest-frame lag (median of the 1000 recovered lags, with the 16th and 84th percentiles as uncertainties) compared to 
the preset $R$-$L$ relation of Eq.~\ref{eq:est_lag_rm_scout}.
The top right panel shows the distribution of the recovered
bias (defined in Eq.~\ref{eq:bias}) analogous to Fig.~\ref{fig:iccf_bias}.
The bottom panel shows an exemplary continuum and line
light curve. Shaded areas show the uncertainty arising 
from the S/N for each light curve, and the data points
show the sampled measurements of the light curve.

The program additionally reports the recovered lag and its uncertainties, the success rate (the fraction of realisations yielding a lag), and the outlier fraction.

We emphasise that the outputs of this program
should be interpreted statistically. The shown lag
on the $R$-$L$ plane with uncertainties represents
the expected distribution of recovered lags for repeated
realisations of the same observing conditions, but not
the simulated recovered lag for a single RM observation.
The bias histogram thus represents the probability distribution of the bias for a single realisation.
A narrow distribution centred on $b=0$ indicates
a robust observing strategy such that a single observation
is likely to produce an unbiased result, whereas a broad
distribution implies that a single light curve observation is 
likely to yield a substantially biased result. Similarly, the success rate 
should not be interpreted as a stand-alone measure of the simulation quality. 
It is defined as the fraction of realisations for which the ICCF identifies a 
sufficiently significant correlation, but it does not quantify the accuracy of 
the recovered lag itself. A high success rate can still be accompanied by a 
biased lag distribution if the majority of successful recoveries converge on an 
incorrect correlation peak. Conversely, a low success rate does not necessarily 
imply poor performance if only the realisations that recover the true variability 
pattern are accepted. The outlier fraction must likewise be interpreted together 
with the success rate. Since it is calculated only from successful lag recoveries, 
a small number of incorrect measurements can produce a large outlier fraction when 
the success rate is low. Conversely, when many realisations are classified as 
successful, the relative outlier fraction may decrease even if the absolute number 
of incorrect lag recoveries increases. For this reason, the success rate, outlier 
fraction, and recovered lag distribution should always be considered together when 
assessing the expected performance of an observing strategy.

\begin{figure*}
    \centering
    \includegraphics[width=1\linewidth]{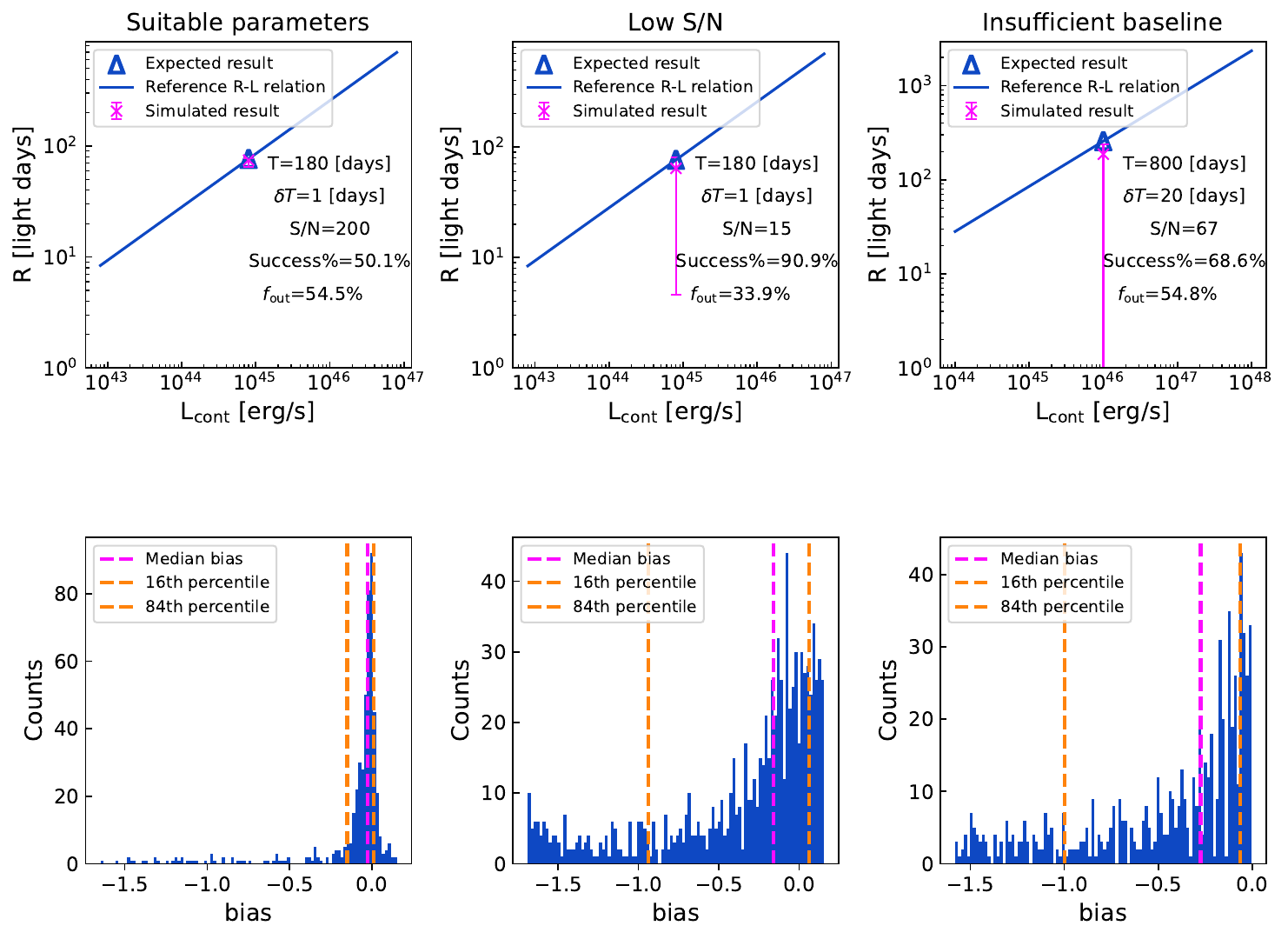}
    \caption{Comparison of the recovery
    expectation for an AGN with $L_{\rm cont}
    =8\times10^{44}~\rm erg~s^{-1}$ at $z = 0.01$
    using different observational noise properties (\textit{left} and \textit{center})
    and one AGN with $L_{\rm cont}
    =1\times10^{46}~\rm erg~s^{-1}$ at $z = 0.5$ (\textit{right}).
    \textit{Upper panels}: Recovered lag
    on the $R$-$L$ plane compared to the
    expected lag with
    the success rate and outlier fraction $f_{\rm out}$. 
    \textit{Lower panels}: Bias histogram of the lag recovery. 
    \textit{Left column}: Recovery using suitable observing
    conditions. \textit{Middle column}: Effect of poor S/N on the lag recovery.
    \textit{Right column}: Effect of an insufficient baseline on the
    lag recovery.}
    \label{fig:rm_scout_comparison}
\end{figure*}

To illustrate the importance of an appropriate
observational setup for reverberation mapping
campaigns, we compare \texttt{scout} predictions for
the same AGN with $L_{\rm cont} = 8\cdot10^{44}
\rm~erg~s^{-1}$ at $z = 0.01$ with two different qualities
of observation and one AGN with $L_{\rm cont} = 1\cdot10^{46}
\rm~erg~s^{-1}$ at $z = 0.5$ with a poor observing strategy in
Fig.~\ref{fig:rm_scout_comparison}.

The left column represents the reference 
configuration with suitable observing conditions.
The upper panel shows the expected position on
the $R$-$L$ relation, while the lower panel 
displays the corresponding bias distribution.
The narrow distribution centred on $b=0$ 
indicates a robust and unbiased lag recovery.

The middle column demonstrates the impact of a
low S/N. In this case, the
median recovered lag is biased by $\sim20\%$ and the distribution is
substantially broadened. 
The apparently high success rate ($\sim$93\%) indicates that the ICCF frequently misidentifies correlated noise as intrinsic AGN variability, resulting in a broad and unreliable recovery.

The right column illustrates the effect of an 
insufficient observing baseline. Although the
ICCF search range still contains the true lag,
the bias distribution is significantly
broadened and its median is biased by 
approximately $25\%$. Consequently, the
predicted outlier fraction, $f_{\rm out}$, is
high, implying that a reliable lag measurement
for a single observed light curve is unlikely.

Overall, these examples demonstrate that
successful lag recovery requires not only an 
unbiased median but also a narrow bias 
distribution centered on $b=0$. Broad bias 
distributions indicate large statistical 
uncertainties and correspondingly low confidence
in the recovered lag from a single reverberation
light curve.

\section{Single epoch mass estimator comparison}
\label{app:mass_comp}

\begin{figure*}
    \centering
    \includegraphics[width=0.22\linewidth]{figures/mass_comparison_bentz.pdf}
    \includegraphics[width=0.22\linewidth]{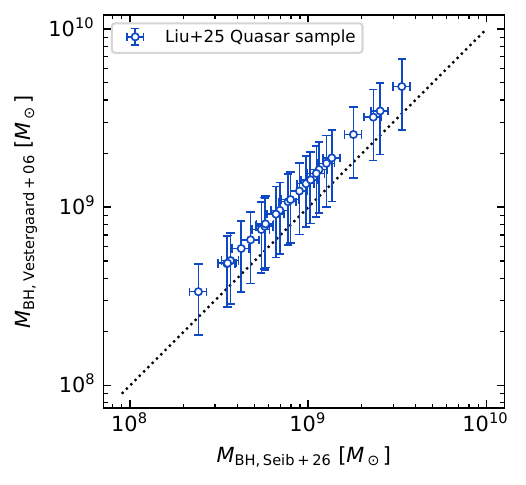}
    \includegraphics[width=0.22\linewidth]{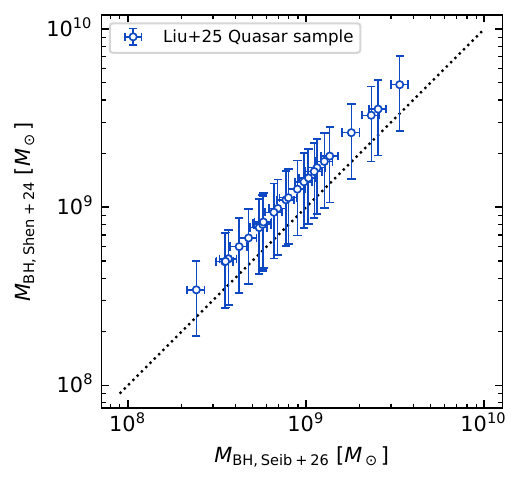}
    \includegraphics[width=0.22\linewidth]{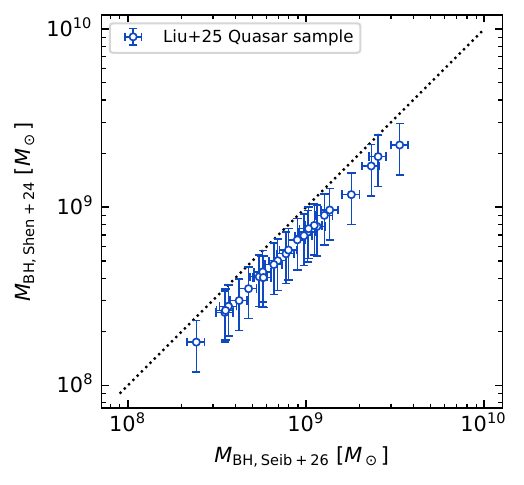}
    \caption{Comparison of black hole masses for the high-$z$ quasar sample of \citet{Liu2025}, inferred with literature SE prescriptions and with this work's estimator.
            \textit{Left to right}: (1) SE mass estimator using
            \citet{Bentz2013} calibrated $R$-$L$ relation; (2) SE mass estimator
            fitted by \citet{Vestergaard2006}; (3) SE mass estimator fitted by
            \citet{Shen2024} assuming a population average virial coefficient and fixed luminosity and line-width slope; (4) SE mass estimator calculated from \citet{Shen2024}
            calibrated $R$-$L$ relation. Uncertainties reflect the intrinsic scatter of each calibration, which dominates over parameter and measurement uncertainties.}
    \label{fig:mass_comp_matrix}
\end{figure*}

As briefly discussed in Sect.~\ref{sec:highz}, we compared the H$\beta$ SE black hole 
mass estimator derived in this work (Eq.~\ref{eq:hb_se}) with several commonly used prescriptions 
from the literature. Figure~\ref{fig:mass_comp_matrix} presents the inferred black hole masses for 
the high-redshift quasar sample of \citet{Liu2025} using our calibration together with those of 
\citet{Bentz2013}, \citet{Shen2024}, and \citet{Vestergaard2006}. For \citet{Shen2024}, we additionally 
compared both their empirically calibrated SE mass estimator and the estimator obtained by directly 
propagating their fitted $R$-$L$ relation. The recipes differ because, in the case of the direct calibration, the calibration masses were derived using a population average virial coefficient and only the offset was fitted as a free parameter. The propagated estimator was derived with our approach, without prescribing a virial coefficient and only fixing the slope of the velocity indicator at $2$.

The inferred black hole masses differ systematically between the various prescriptions, demonstrating the 
strong dependence of SE mass estimates on the adopted calibration. These differences primarily originate 
from variations in the underlying $R$-$L$ relation, but are further amplified by assumptions regarding the 
virial coefficient, the adopted line-width definition, and the calibration methodology.

The calibration presented in this work is based on the largest bias-corrected reverberation mapping sample 
currently available and yields one of the smallest intrinsic scatters among existing H$\beta$ $R$-$L$ relations. 
Consequently, it provides a statistically well-constrained calibration while relying on comparatively few 
assumptions. In contrast, the prescriptions of \citet{Shen2024} and \citet{Vestergaard2006} are calibrated 
directly against reverberation-mapped black hole masses and therefore inherit assumptions about the 
population-averaged virial factor and the adopted line-width measure. Furthermore, the \citet{Shen2024} calibration 
fixes the luminosity exponent in addition to the commonly fixed line-width exponent, fitting only the normalisation.

Overall, the comparison illustrates that systematic differences between commonly used SE mass estimators can exceed 
their formal statistical uncertainties. The choice of calibration therefore represents a significant source of 
systematic uncertainty when comparing black hole masses across different studies. In this context, the calibration 
presented here offers the advantage of being derived from a larger and more homogeneous reverberation mapping sample 
while minimising additional assumptions beyond the empirical $R$-$L$ relation.

\end{appendix}

\end{document}